%% file: main.tex
\documentclass[numbers,compress,3p,times,review]{elsarticle} 
\usepackage[utf8]{inputenc}
\usepackage{natbib}
\setcitestyle{square, comma, numbers,sort&compress, super}
\usepackage{natbib}
\usepackage{lmodern}
\usepackage[dvipsnames]{xcolor}
\usepackage{booktabs}
\usepackage{multirow}
\usepackage{graphicx}
\usepackage{subcaption}
\usepackage{textcomp, gensymb}
\usepackage{gensymb}
\usepackage{siunitx}
\usepackage{amsmath,amsfonts,amssymb}
\usepackage{epsfig}
\usepackage{cleveref}
\usepackage{amsmath}
\usepackage{amssymb}
\usepackage{algorithm}
\usepackage{algpseudocode}
\usepackage{enumitem}

\title{Layer-to-layer Closed-loop Switched Heating and Cooling Control of \\
the Laser Powder Bed Fusion Process\tnoteref{tnote1}}

\journal{Additive Manufacturing}

\author[inst1]{Bar{\i}\c{s} Kavas\corref{cor1}}
\ead{bkavas@ethz.ch}
\author[inst2,inst3]{Efe C. Balta}
\ead{efe.balta@inspire.ch}
\author[inst1]{Lars Witte}
\ead{lawitte@ethz.ch}
\author[inst1]{Michael R. Tucker}
\ead{mtucker@ethz.ch}
\author[inst3]{John Lygeros}
\ead{jlygeros@ethz.ch}
\author[inst1]{Markus Bambach}
\ead{mbambach@ethz.ch}

\cortext[cor1]{Corresponding author}

\address[inst1]{Department of Mechanical and Process Engineering,
ETH Z\"urich, Z\"urich 8092, Switzerland}
\address[inst2]{Control and Automation Group, inspire AG, Zürich 8005, Switzerland}
\address[inst3]{Department of Information Technology and Electrical Engineering, Automatic Control Laboratory, ETH Zürich, Zürich 8092, Switzerland}

\tnotetext[tnote1]{This manuscript has been submitted to \textit{Additive Manufacturing} and is currently under review.}

\newdefinition{rmk}{Remark}

\begin{document}



\input{0-abstract}

\maketitle

\input{1-introduction}
\input{2-Theory}
\input{3-Material_and_Methods}
\input{4-results_discussion}

\input{5-discussion}
\input{6-conclusion}

\section*{Declaration of Generative AI and AI-assisted Technologies in the Preparation of this Manuscript}

During the preparation of this work, the authors used AI-based tools to support text refinement and clarity. 
After using this tool, the authors carefully reviewed and edited the content as needed and take full responsibility for the final version of the manuscript.

\section*{Acknowledgemnets}

The authors would like to thank Julius Valentin Van der Kuip for his support with specimen preparation and metallographic characterization.



\bibliographystyle{model1-num-names}

\bibliography{cas-refs}

\end{document}

%% file: 0-abstract.tex
\begin{abstract}

This study investigates the stabilization of interlayer temperature in the laser powder bed fusion process through a novel switched layer-to-layer closed-loop feedback controller. 
The controller architecture aims to measure the interlayer temperature by a laterally positioned thermal camera and maintain a preset reference temperature by switching between the heating mode through dynamic laser power adjustment and the cooling mode by assigning interlayer dwell time to allow cooling between layers.
The switching controller employs a feedback optimization control algorithm for the heating mode to adjust the laser power, and a triggering algorithm that increases the interlayer dwell time until the interlayer temperature reaches the reference value.
Additionally, the study compares the performance of the proposed controller in both supported and unsupported overhanging parts to evaluate the effect of support structures on the controller performance as well as the thermal behavior of overhanging parts.
Key results demonstrate the controller's effectiveness in stabilizing interlayer temperature across varying cross-sectional areas while remaining within the material's stable processing zone. 
In the heating mode, the controller efficiently tracks the reference temperature, even in geometries with significant cross-section variation. 
During cooling, the controller adjusts dwell times to enhance thermal control in overhanging sections.
The controller's robustness is further validated by its performance with unsupported parts, where the overheating effect is more pronounced, and in supported parts, where thermal conduction to the build plate is enhanced. 
The study also identifies trade-offs among process efficiency, energy consumption, and build time. 
Supported parts exhibit reduced overheating but consume more energy and material, while unsupported parts stabilize interlayer temperature faster but with longer build times due to increased dwell time assignments. This tradeoff is more than compensated by a reduction in post-processing effort.
The research highlights notable improvements in interlayer temperature control for geometries prone to excessive thermal stresses. 
Moreover, the introduction of interlayer dwell time offers a practical solution to maintaining thermal stability in complex geometries. 

\end{abstract}

%% file: 1-introduction.tex
\section{Introduction}
\label{sec:Introduction}
Laser powder bed fusion of metals (PBF-LB/M) is a critical manufacturing technology utilized in various high-value addition industries including the healthcare, aerospace, and automotive sectors~\cite{Ngo2018} it enables cost-effective production of high-performance parts with intricate designs \cite{Craeghs2011}. 
Among other factors, the precision and quality of the process are fundamentally dependent on the positional and movement accuracy of the scanner, the actual laser power applied, and the optical path design of the beam~\cite{Kaufmann2016}. 
Due to the interplay of various factors such as recoating, and laser power application, ensuring the desired quality of printed parts poses significant challenges for the robustness of the PBF-LB/M process~\cite{stavropoulos2021robust}.
Additionally, process conditions and the quality of parts are greatly influenced by their geometry, due to the changing heating and cooling conditions on wide-ranging time and spatial scales~\cite{kozak2018accuracy}.

A key parameter affecting the quality of the PBF-LB/M process is the \textit{inter-layer temperature} (ILT), which describes the spatially averaged temperature of the exposed surface of a particular part or volume within a layer interval in a static time with respect to the last laser exposure.
ILT, which is influenced by the thermal history of the part, its geometry, and material properties, is a \emph{global} thermal phenomenon that fluctuates temporally on the scale of tens-of-seconds to minutes through the layers. 
The primary determinant of ILT is the balance between the energy input by the scanning laser and its dissipation as the part cools down during the exposure of other parts and the recoating process. 
Cooling primarily occurs through conduction to the solid mass underneath the surface (into the build platform through previously printed layers).
To a lesser extent, heat conduction occurs through the surrounding powder, along with heat loss due to convection and radiation~\cite{Mugwagwa2016}.
The energy input per layer is proportional to the exposed top surface area and the thermal conductivity is proportional to the area and length of the previously solidified material~\cite{Patterson2017}.
Under steady-state processing conditions and static energy input, upward facing part features (converging geometries with decreasing cross-sectional areas) tend to have a decreasing ILT, vertical features with constant cross-sectional areas will have a constant ILT, and overhanging features (divergent geometries with increasing cross-sectional areas) tend to have an increasing ILT~\cite{sames2016metallurgy, LopezTaborda2021}. 
In the last case, both the total heat input and thermal resistance increase over time, resulting in increasing ILT in subsequent layers.

Variation in ILT during a build job has been linked to various process anomalies and quality concerns~\cite{sames2016metallurgy,debroy2018additive}. 
On the melt pool scale, the temperature of the surface just before laser exposure is known as the \textit{pre-heat temperature}, which includes the pre-heating effects from adjacent weld beads within a hatch field. 
ILT significantly affects the pre-heat temperature, and its impact on solidification dynamics has been extensively studied~\cite{park2022effect}. 
Thermal gradients can lead to thermal stress, which in turn can cause mechanical distortion~\cite{afazov2017distortion}. 
If this distortion exceeds the layer thickness in the build direction, it may result in recoater interaction. 
The correlation between ILT and residual stress accumulation is well-documented~\cite{megahed2016metal, anandan2019distortion}. 
Research by Yavari et al. and Kobir et al.~\cite{yavari2021part, kobir2022prediction} indicates that changes in ILT can result in porosity and micro-structural variation, as well as recoater interactions. Likewise, Nahr et al. demonstrated a relationship between ILT and microstructure in Ti6Al4V and MS1 (CITE). 
Recent work by Kusano et al. demonstrated a link between the ILT and cracking in IN738LC, which may be controlled through modeling (CITE1) and closed-loop control (CITE2).
Moreover, Li et al. \cite{li2021multi} validated the direct relationship between the rate of ILT increase and grain size increase, along with reduced micro-hardness in titanium aluminide alloys. 
As such, varying ILT is considered to be the primary factor that limits printing of parts with rapidly changing cross-sectional areas in the build plane~\cite{Savalani2016, Papadakis2018, Motibane2019}.

\subsection{ILT stabilization strategies}
\label{sec:strategies}
There are two main approaches in the literature for stabilizing the ILT: 1. reorienting the part and adding supports, and 2: adapting the rate and location of energy input through open- or closed-loop process control.

\subsubsection{Reorienting and adding supports}

By reorienting the part, one can find an optimal orientation that minimizes the variation in cross-section and the overhang angle~\cite{sames2016metallurgy}. 
However, this is not always practical in light of other considerations, such as part symmetry, surface finish requirements, or the desired grain growth direction.
Support structures can also be added to the part, which has three main effects.
First, they provide additional conductive heat flow paths to equalize the temperature distribution along the build direction resulting in reduced thermal stresses~\cite{paggi2022implementation}.
Another effect is the increased layer build time which slows down the process and allows heat to dissipate.
Finally, support structures provide mechanical resistance against distortion and warping caused by thermal gradients \cite{khobzi2022role}.
Support structures have been commonly applied to especially steep overhanging areas for introducing thermal and mechanical stabilization of the build~\cite{jiang2018support}.
Therefore, although adding additional supports can be advantageous, it also introduces trade-offs, including increased total build time and cost, as well as the need for an additional step to remove them during post-processing.
Though still challenging to model from a thermal perspective, the literature is abundant with the design optimization and effect of the support structures ~\cite{javidrad2023review,dimopoulos2023support}.
However, there are no studies that evaluate the effect of support structures on the process and temperature control.

\subsubsection{Open- and closed-loop process control in PBF-LB/M}
\label{sec:open_close_loop}

Applying process control is an approach for improving thermal stability by adjusting the energy input.
Parameter optimization studies typically use a canonical part design to remove the effect of geometry over the quality criteria and isolate the effects of process variables.
However, standardized geometries fail to represent the geometrical complexity of real-world parts~\cite{cao2021optimization}.
Therefore, adjusting one or more of the processing parameters based on the ILT of the part has been a recent interest in the PBF-LB/M closed-loop control literature.
Fundamentally, there are two main methods for the PBF-LB/M process control: feed-forward and feedback control.
Feed-forward control depends on a high-fidelity model or pre-recorded data that captures the process behavior sufficiently to specify process inputs that stabilize the process in advance~\cite{riensche2022feedforward,kusano2024effects}. 
While somewhat effective, its major shortcoming is the inability to account for unmodeled process perturbations and the reliance on known material, machine, and process parameters.
By contrast, feedback control uses in-situ process measurement data to adjust the process parameters.
This enables the controller to stabilize the process despite unmodeled disturbances.

Closed-loop controllers gained significant popularity in the literature in recent years since they often do not need computationally expensive models to represent process behavior for calculating real-time control actions \cite{spector2018passivity,asadi2021gaussian}. 
Furthermore, process perturbations can be minimized by carefully selecting the measurement signal and control strategy\cite{liao2022layer,zuliani2022batch}.

Closed-loop feedback control strategies are commonly studied in two different time and spatial scales: within-layer control and layer-to-layer control.
Numerous studies focused on developing within-layer control strategies~\cite{benda1994temperature,kruth2007line, kruth2007feedback, craeghs2010feedback, Renken2017, Renken2018, Renken2019, hussain2021feedback} to stabilize the meltpool temperature despite the within-layer preheat temperature variations within the confines of \emph{local} thermal conditions. 
In contrast, layer-to-layer control loops address \emph{global} thermal conditions that arise as ILT varies across layers~\cite{liao2023layer}. 
Among the few studies, Vasileska et al.~\cite{Vasileska2020} developed a layer-to-layer strategy using a high-speed CMOS camera to measure melt pool size variations and modulated the laser pulse frequency to adjust the energy input. 
Their follow-up study applied the same control architecture to an overhanging bridge-like structure, addressing overheating due to reduced conduction ~\cite{vasileska2022novel}. 
Both studies compensated for residual heat due to geometrical variations by adjusting energy input in subsequent layers, thus stabilizing within-layer temperature variation without addressing layer-to-layer heat accumulation or ILT changes. 
Nahr et al. demonstrated a similar approach based on optical tomography \cite{nahr2025advanced} for processing Ti6Al4V and MS1. 
The processed thermal image was overlaid on the scan vectors of the next layer.
These were subdivided into microvectors, each with a power value adapted inversely proportional to the thermal emissions. 
Using this approach, together with a minimum $40~s$ interlayer time for MS1, the ILT was stabilized throughout the build of some overhanging cone structures, resulting in distinct microstructures relative to the uncontrolled case.

The recent study by the authors~\cite{kavas2023layer} proposed a layer-to-layer control approach to stabilize ILT to a reference value by controlling the laser power input for each layer.
Although this study stabilized the ILT for various overhang angles, the part geometry eventually resulted in uncontrolled overheating that drove the controlled laser power out of the process window and resulted in the lack-of-fusion porosity.
Furthermore, surface condition-related non-linearities in the thermal camera measurements caused by the porosity that negatively affected the controller performance were observed.
Maintaining the ILT by only adjusting the laser parameters will eventually force the controller out of the stable limit of the process window.
Most of the referenced studies use laser power as the energy input parameter for actuation to stabilize ILT.
Due to the limit posed by the process window to obtain a high-quality part, actuating the power input, consequently, the amount of heat input to every layer, was shown to be insufficient to stabilize ILT without introducing porosity.

The term \emph{interlayer dwell time (IDT)} is introduced to describe an artificial waiting time that is added to the recoating delay time to increase the cooling period to conduct heat away from the part's surface.
IDT has been studied extensively for direct energy deposition (DED) modalities, as described by Mohr et al.~\cite{mohr2020effects}.
Kusano et al. demonstrated a feedforward model-based approach to add IDT to control ILT in PBF-LB/M \cite{kusano2024effects}, which they later extended to include thermal image feedback control \cite{kusano2024controlling}.
Both of these studies focused solely on cooling without attention to heating control, which was sufficient to prevent cracking in IN738LC.
In Nahr et al.'s study on layer-to-layer heating control the authors mention that it was necessary to add $40~s$ of static IDT to prevent overheating and process instabilities for the MS1 specimens \cite{nahr2025advanced}.

Krauss~\cite{krauss2017qualitatssicherung} identified three types of thermal cycling in PBF-LB/M that influence effective cooling rates and melt pool dimensions, as referenced by Kruth et al.~\cite{kruth2007consolidation}: hatch cycling, scan field cycling, and layer cycling. 
The first and last types of thermal cycling were found to be critical, underlining the importance of the layer-to-layer thermal cycles.
Xu et al.~\cite{xu2017situ} considered IDT as a variable for powder bed-based AM and conducted experiments on PBF-LB/M of Ti-6Al-4V. 
They created cylindrical specimens with diameters ranging from 0.8 mm to 12 mm and a height of 30 mm using four different IDT assignments (1 s, 5 s, 8 s, and 10 s). 
Longer IDT led to finer grains, resulting in higher yield strength and ultimate tensile strength. 
Lui et al.~\cite{lui2017new} found that IDT affected the width of alpha-laths in PBF-LB/M experiments with Ti-6Al-4V, with longer ILTs producing finer laths. 
They noted that an IDT of 8 s is a threshold for this refining effect, and recoating times should be added to meet the IDT definition used in this study.
~\cite{mohr2020effects} emphasizes the effect of shorter IDT assignments by showing increased grain size, melt pool depth, and decreased hardness.
Recently, Olleak et al.~\cite{olleak2024understanding} experimentally investigated two different build plans with and without additional IDT for an overhanging structure with Ti-6Al-4V.
Their results indicate significant differences in the preheat temperatures, cooling rate, and ILT.
Furthermore, they observed martensitic transformation due to high cooling rates and varying hardness values in the microstructure, which strongly indicates the effect of IDT has on the material properties of printed parts.

The natural variation of IDT due to factors such as build layout, scan strategies, and part geometry has only recently begun to be studied and optimized for process control.
Riensche et al.~\cite{riensche2022feedforward} applied a feed-forward approach to optimize the laser power and IDT to stabilize ILT across the layers supported by a physics-based simulation.  
Their results show that feed-forward control produced parts with finer grain size, increased microhardness, improved geometric integrity and resolution, and reduced surface flaws.
Their study represents the only example in the literature that incorporates IDT and energy input together for enhanced ILT stabilization, however, only by a feed-forward parameter assignment.
Behrens et al.~\cite{behrens2024temperature} proposed an algorithm to calculate adaptive IDT for thermal stability aided by a part-scale temperature simulation.
Their feed-forward approach was implemented in a simulation study, which showed improved ILT stability in parts with overhang regions and geometric features that cause heat buildup.
Kusano et al.~\cite{kusano2024effects} applied a part-scale finite element thermal analysis to predict the ILT under varying IDT conditions.
This was experimentally validated, showing a direct influence on microstructure, hardness, and cracking behavior.
Their follow-on study~\cite{kusano2024controlling} briefly describes the use of a thermal camera to realize closed-loop control by varying IDT, but very few details are provided regarding the implementation.
More recently,~\cite{kavas2025physics} implemented a 1D finite-volume model to measure the temperature difference between two surfaces of a cantilever beam part to feed-forward assign IDT per layer.
Their experiments achieved \~20\% less distortion at the tip of the cantilever beam by applying the same amount of total IDT per layer assignment.
Overall, the existing literature on IDT actuation for ILT stabilization in PBF-LB/M relies on computationally complex process models with feed-forward parameter settings that would fail to capture process deviations and perturbations.

\subsection{Scope of the Study}
\label{sec:scope_of_study}

The existing literature highlights several research gaps related to the closed-loop control of the PBF-LB/M process for ILT stabilization.
First of all, the majority of existing closed-loop control approaches for the PBF-LB/M process are based on regulating the heating mechanism by adjusting the energy input.
While IDT control has only very recently become a topic of interest, only one has applied it in a closed-loop manner \cite{kusano2024controlling}.
No studies have been identified that combine control of the heating (i.e., process energy inputs) with cooling (i.e., energy dissipation time) within a unified closed-loop feedback control for this process.
Second, most process control studies focus on simple overhanging geometries where the only tendency is to overheat.
Only one study looked at converging and diverging geometries that attempted to track a prescribed ILT despite net heat accumulation and dissipation \cite{nahr2025advanced}.
They found it necessary to add a relatively long static waiting time in order to prevent accumulation, which is a practical solution at the significant expense of increasing the overall build time.
Third, although the controllability limits for varying overheating trends are investigated, the effect of adding support structures on the ILT stabilization performance of a layer-to-layer control has not been explicitly studied.
Besides these contributions, this study also provides a quantitative view of the trade-offs between ILT stabilization, build time, and energy use, placing the results in the broader context of process efficiency and sustainability in PBF-LB/M.

Considering the identified research gaps, the main study and contributions of this paper are:
\begin{itemize}
    \item A novel layer-to-layer closed-loop feedback control architecture that seamlessly switches between cooling (IDT modulation) and heating (power input) mechanisms during the process that is capable of indefinitely stabilizing ILT in the PBF-LB/M process with experimental validation on diverging and converging geometries with minimal additional processing time
    \item Evaluation of the effect of including support structures on the controller performance for overhanging geometries
    \item An analytical approach for estimating the ILT and resulting IDT for cooling to a reference temperature based on geometry
\end{itemize}

The control architecture and its implementation are provided in \Cref{sec:theory}. 
\Cref{sec:materials_methods} describes the experimental procedure with detailed descriptions of the characterization methods, hardware, and software used to reproduce the procedures described. 
Experimental measurements are shared in~\Cref{sec:results} and interpreted in detail in \Cref{sec:discussion}.
Concluding remarks and future directions are given in \Cref{sec:conclusion}.

%% file: 2-Theory.tex
\section{Theory}
\label{sec:theory}

This section analyses the source of ILT change and the within-layer cooling of the exposure surface, along with the interplay between them.
For this, a simple 1D model is derived for the relevant thermal dynamics for the layer-to-layer PBF-LB/M process with converging and diverging cross-sectioned geometries and to describe the amount of conductive cooling rate change based on the layer and the part geometry. 
Subsequently, the proposed closed-loop layer-to-layer switch heating and cooling control architecture is explained in detail with the heating mode evaluating the overheating while the cooling mode actuates the amount of time needed to reach a reference ILT.

\subsection{Layer-to-layer heat buildup}
\label{sec:heat_buildup}

After each layer is printed, the build plate descends by one layer thickness, and a fresh layer of powder is spread via the recoating process. 
During this period, heat input by the laser to the newly solidified material is transferred to the surroundings.

The following relation thus represents the ILT:

\begin{equation}\label{eq:layerOp}
    T_{k+1}(0) = \mathcal{S}_k(T_{k}(\tau_e(k)),(\tau_c+\tau_{idt}(k))),
\end{equation}

where $\tau_e$ is the laser exposure time, $\tau_c$ is the recoating time, $\tau_{idt}$ is the IDT, and the subscript $k = 1,\ldots, n_{\ell}$ denotes the layer index for a process with a total of $n_{\ell}$ layers.
$\mathcal{S}$ denotes layer-to-layer process thermal dynamics, and $T_{k+1}(0)$ is the homogeneous temperature of the part's surface immediately before the exposure of the next layer and thus corresponds to the ILT, which is aimed to be stabilized.
Given the defined time terms, total layer processing time $\tau_{\ell}$ for layer $k$ is described as

\begin{equation}\label{eq:idt}
    \tau_{\ell}(k) = \tau_{e}(k) + \tau_{c} + \tau_{idt}(k).
\end{equation}

where the recoating time $\tau_{c}$ is constant due to process- and platform-specific constraints, and exposure time $\tau_{e}$ and IDT $\tau_{idt}$ are layer dependent.

The unforced cooling function is defined as:
\begin{equation}\label{eq:cooling}
    f_k(T_{k}(t)) =  \mathcal{K}_k\left( T_{k}(t) - T_s \right) + h_{co} a_k\left( T_e - T_{k}(t) \right) = A_k T_{k}(t) + v_k,
\end{equation}
where $a_k$ is the surface area at layer $k$, $h_{co}$ is the convection coefficient, $T_s$ is the substrate temperature, $T_e$ is the ambient temperature, and $\mathcal{K}_k$ denotes the equivalent thermal conductance to the previously consolidated layers at layer $k$.
Evidently, the unforced cooling function $f_k(\cdot)$ captures the passive conductive and convective heat transfer that governs the evolution of the surface temperature between successive layers.
This description assigns the conduction-related cooling effect to the term $\mathcal{K}_k$, therefore, its exact value depends on the existing temperature profile under the surface. 

Since~\eqref{eq:cooling} is affine in T, it can be compactly written as $A_k = \mathcal{K}_k - h_{co}a_k$ (i.e., rearranged conductance term) and $v_k =  -\mathcal{K}_k T_s + h_{co}a_k T_e$ (i.e., cooling term).
Neglecting changes in boundary conditions caused by the deposition of new material during the recoating process, the thermal dynamics of the part surface can be approximated by the following expression as described by \cite{liao2022layer}:

\begin{equation}
\label{eq:layer_dyn_temp} 
c_p \rho \Delta l a_k \dot{T}_k(t) = 
\begin{cases} 
f_k(T_k(t)) + \eta \mu_k(t), & \text{if } t \in [0, \tau_k] \\
f_k(T_k(t)), & \text{if } t \in (\tau_k, \tau_k + \tau_c + \tau_{idt}]
\end{cases}
\end{equation}

was derived to describe the layer-to-layer heat build-up behavior, where $c_p$ is the specific heat capacity of the material, $\rho$ is its density, and $\Delta l$ is the reference element height, $\tau_{\ell}(k)$ is the layer-dependent execution time, $\eta$ is the absorption coefficient, and $\mu_k$ is the laser power.


Unlike the laser power, the cooling duration can be determined using in-situ temperature measurements after recoating, without requiring future state predictions. This availability is also exploited in the cooling controller mode design as described in \Cref{sec:controller_architecture}.  
However, even in the absence of closed-loop control, the ability to estimate the $\tau_{idt}$ required to reach a target reference temperature based solely on the geometry offers valuable means to mitigate the effects of overheating and thermal accumulation, as indicated by previous studies \cite{kusano2024effects}.

To this end, the thermal model introduced in Equation~\eqref{eq:layer_dyn_temp} is used to approximate the temperature evolution during the cooling phase and, ultimately, to predict the required $\tau_{idt}$. 
The model estimates the ILT by accounting for the layer-wise heating and cooling of the surface, while introducing the equivalent thermal conductance $\mathcal{K}_k$ to capture both the influence of the underlying geometry and the accumulated thermal history on conductive cooling over a single element.
The accuracy of the proposed approach, therefore, relies critically on assigning an appropriate value to $\mathcal{K}_k$ such that it reflects the cooling characteristics of each layer.

Two alternative fitting strategies are proposed for the effective thermal conductance $\mathcal{K}_k$:
The first approach models $\mathcal{K}_k$ as a function of the layer's cross-sectional area. 
The relationship is expressed as

\begin{equation}
\hat{\mathcal{K}}_k(A_i) = \frac{\alpha}{A_i} + \beta,
\end{equation}

where \( A_i \) denotes the cross-sectional area of the \( i \)-th layer, and \( \alpha \), \( \beta \) are scalar parameters to be identified. 
Given a dataset of inversely identified (calculated based on known measurements of ILTs) $\mathcal{K}_k$ values \( \{(A_i,\mathcal{K}_{k,i})\}_{i=1}^{N} \), the fitting is performed by minimizing the squared residuals using the Levenberg-Marquardt algorithm:

\begin{equation} \label{eq:inversely_proportional}
\min_{\alpha, \beta} \sum_{i=1}^{N} \left( \mathcal{K}_{k,i} - \left( \frac{\alpha}{A_i} + \beta \right) \right)^2.
\end{equation}

This model captures the empirical observation that the range of layers with larger cross-sectional areas exhibits higher effective thermal conductance due to enhanced heat dissipation.

As a baseline comparison, a second approach assumes a constant thermal conductance parameter across all layers:

\begin{equation}
\hat{\mathcal{K}_k} = \gamma,
\end{equation}

where \( \gamma \) is a scalar value fitted to minimize the total squared error over the same dataset:

\begin{equation}\label{eq:constant}
\min_{\gamma} \sum_{i=1}^{N} \left( \mathcal{K}_{k,i} - \gamma \right)^2.
\end{equation}

The accuracy of each fitting strategy is evaluated by comparing the predicted inter-layer temperatures to experimental measurements. 
As a quantitative metric, the Mean Absolute Percentage Error (MAPE) was used:

\begin{equation}
\text{MAPE} = \frac{100}{N} \sum_{i=1}^{N} \left| \frac{T_i - \hat{T}_i}{T_i} \right|,
\end{equation}

where \( T_i \) is the measured temperature and \( \hat{T}_i \) is the predicted value, both are calculated in Kelvin units.
The values used for the prediction calculation in this study for the stainless steel 316L (1.4404) used in the experiments are given in \Cref{tab:simulation_parameters}.

\begin{table}[h!]
\centering
\caption{Simulation Parameters for stainless steel 316L and the experimental setup}
\label{tab:simulation_parameters}
\begin{tabular}{@{}lll@{}}
\toprule
\textbf{Parameter} & \textbf{Value} & \textbf{Unit} \\
\midrule
Specific heat capacity, $c_p$               & 468.0      & \si{\joule\per\kilogram\per\kelvin} \\
Density, $\rho$                    & 8000.0     & \si{\kilogram\per\cubic\meter} \\
Convection coefficient, $h_{co}$   & 15.0       & \si{\watt\per\square\meter\per\kelvin} \\
Absorptivity, $\eta$               & 0.4        & -- \\
Substrate temperature, $T_s$       & 473.15     & \si{\kelvin} \\
Ambient temperature, $T_e$         & 293.15     & \si{\kelvin} \\
Element height, $\Delta l$        & 0.0006     & \si{\meter} \\
Initial temperature, $T_0$         & 473.15     & \si{\kelvin} \\
\bottomrule
\end{tabular}
\end{table}


\subsection{Controller architecture}
\label{sec:controller_architecture}
In this work, a controller strategy is proposed to counteract the outlined factors in the previous section that affect the ILT variance across the layers to achieve a stabilized ILT in both converging and diverging cross-sections while remaining within the process window.
The proposed method leverages the competing mechanisms of heating (laser exposure) and cooling (recoating time and IDT) within the natural thermal cycle of a layer as explained in detail in Section \ref{sec:heat_buildup}.

\begin{figure}[ht!]
\begin{center}
\includegraphics[width=0.7\columnwidth]{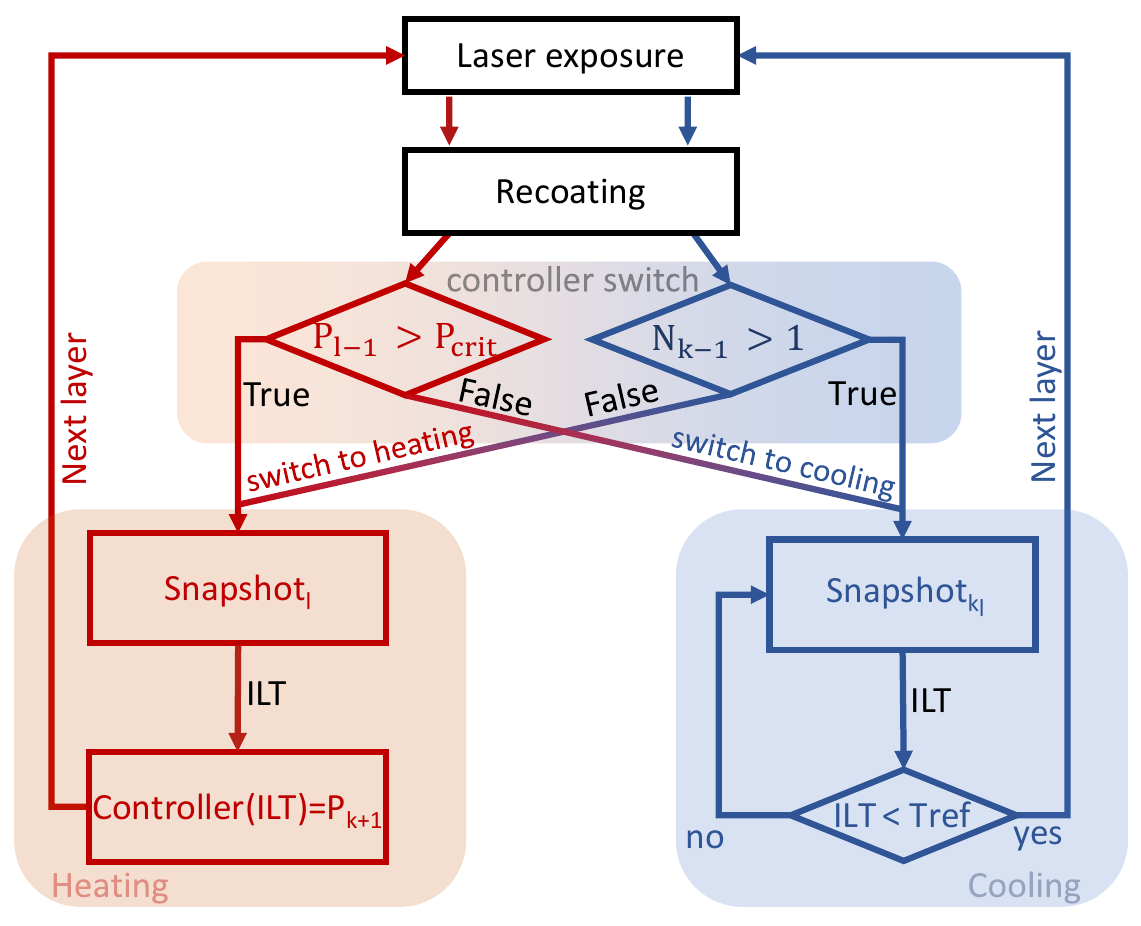}
\caption{The proposed heating and cooling switch controller architecture. The method leverages the energy input actuation capability to control the ILT (heating) until the overheating dominates and the dynamic laser power assignment saturates the lower boundary of the process window. Once reaching the boundary, the cooling trigger is initiated to wait the required amount of time to reach stable ILT.}
\label{fig:proposed_method}
\end{center}
\end{figure}

The proposed control architecture for ILT stabilization is summarized in Figure \ref{fig:proposed_method}. 

The process control flow is composed of 3 main components: the controller switch, heating controller, and cooling controller.

\subsubsection{Controller switching}

The switching algorithm is given in the algorithm environment in \Cref{alg:l2l_controller}.

\begin{algorithm}[H]
\caption{Layer-to-Layer Controller with Heating--Cooling Switch Algorithm}
\label{alg:l2l_controller}
\begin{algorithmic}[1]

\State \textbf{Initialization:}
\State Wait for buffer layers to avoid platform transients
\State Set controller mode $\gets$ \texttt{HEATING}
\State Set reference temperature $T_{ref}$

\For{each layer $l = 1,2,\dots$}
    \State Perform laser exposure and recoating

    \If{controller mode $=$ \texttt{HEATING}}
        \If{$P_{l-1} > P_{low}$}
            \State Apply dynamic power adjustment or saturate at upper power limit
            \State Continue heating control (assume overheating not imminent)
        \Else
            \State Assign $P_{l} \gets P_{low}$
            \State Switch controller mode $\gets$ \texttt{COOLING}
        \EndIf
    \EndIf

    \If{controller mode $=$ \texttt{COOLING}}
        \Repeat
            \State Measure ILT via snapshot
        \Until{ILT $\leq T_{ref}$}
        \If{$N_{l-1} \leq 1$}
            \State Switch controller mode $\gets$ \texttt{HEATING}
        \EndIf
    \EndIf
\EndFor

\end{algorithmic}
\end{algorithm}

The controller is proposed to be initiated after several buffer layers, depending on the build setup, to avoid the initial transitional effects arising from the build platform.
Once the controller is initiated after the buffer layers, the laser exposure and recoating steps are performed in every layer as per the usual process.
The controller is initialized in heating mode since the ILT in the earlier layers of a build is expected to be cooler than the reference temperature. 
However, this assumption is highly dependent on the reference temperature selection.
After recoating, the controller switch logic either decides to stay in the heating mode or switch to the cooling mode.
Under heating control, this decision is made by evaluating the power assignment of the previous layer.
As long as the assigned laser power for the layer is greater than the lower threshold value ($P_{l-1}>P_{low}$), the controller is either dynamically changing the power or the laser power remains saturated at the upper limit.
In both scenarios, the assumption is that overheating is not imminent; thus, the heating control remains active.
Once the power assignment reaches the lower threshold $P_{l-1} \leq P_{low}$, it is assumed that the ILT increase cannot be stabilized through the capacity of dynamic power control (i.e., heating-mode) thus the overheating is imminent.
In this case, the saturation value of $P_{low}$ is assigned, and the controller switches to the cooling-controlled loop.

The cooling-controller loop operates by delaying and triggering the next layer's execution based on the ILT measurement.
The ILT is monitored by periodic snapshots, and the algorithm adds further dwell time if the ILT is larger than the reference value.
The number of snapshots taken during each layer $l$ is described by $n$, and the algorithm evaluates the number of snapshots taken on the previous layer $N_{l-1}$.
Once $N_{l-1}$ is equal or less than one, it indicates that the ILT reached $T_{ref}$ and laser power can again be used to control the ILT.
Therefore, the algorithm switches to the heating-controlled loop.

The key characteristic of the controller switch is operated exclusively in either the cooling- or heating-controlled modes, ensuring that both mechanisms are not activated simultaneously.

\begin{figure}[ht!]
\begin{center}
\includegraphics[width=0.6\columnwidth]{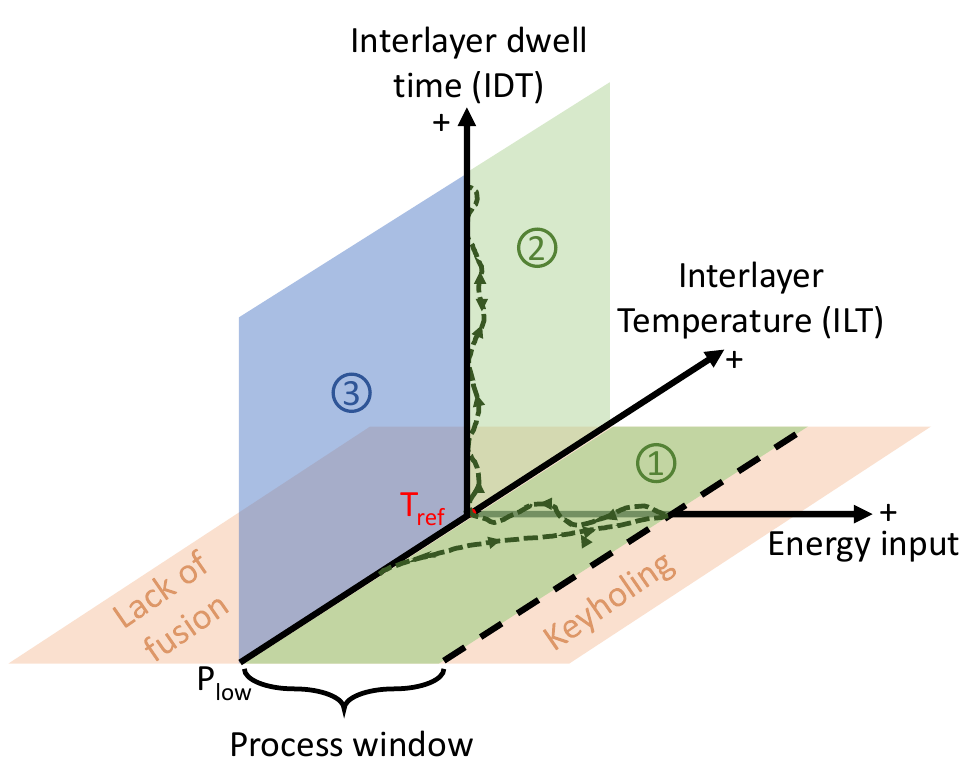}
\caption{Graphical representation of the control inputs as IDT and the energy input relationship to the ILT based on the proposed control algorithm.
The green regions 1 and 2 describe the heating and cooling control loop dynamic control areas, respectively. 
The blue plane marked as 3 is disabled by the algorithm since the cooling mode is only initiated once the ILT is greater than $T_{ref}$.
The red regions define the outside of the process window for the energy input.}
\label{fig:process_map}
\end{center}
\end{figure}

The implications of the proposed controller switch algorithm on the process map are given in~\Cref{fig:process_map}.

The abstract diagram shows the relationship between two inputs of the proposed controller and the resulting ILT.
The energy input vs ILT plane is shown in red while the IDT vs ILT plane is shown in blue.
The process window ($U$) is defined in the energy input axis, holding other process parameters constant.
While insufficient energy results in lack-of-fusion defects, excessive energy initiates keyholing and induces porosity in the microstructure.
The reference temperature is defined on the ILT axis as $T_{ref}$.

\begin{itemize}
    \item Upon initiation of the heating mode, the energy input can be defined anywhere within the process window (region 1) until the settling of the signal $y_k$ around the reference value $T_{ref}$.
    \item If persistent overheating drives the controller to saturate in the lower end of the process window ($P_{l-1} \leq P_{low}$), the controller switches to the cooling controller, and the energy input is fixed at $P_{low}$. 
    The cooling mode can move within region 2, compensating for the ILT trends depending on $\Delta a$.
    On the contrary, the controller without IDT addition (with only heating control) would drive the ILT past the reference value and keep overheating.
    \item Since the controller switch is initiated only when the energy input reaches $P_{low}$, IDT cannot follow any trajectory during $P_{l} \leq P_{low}$, which prohibits the controller from pushing the process into region 3.
\end{itemize}

Note that cooling and heating modes represent competing objectives for ILT stabilization, as elaborated in the previous \Cref{eq:layer_dyn_temp}, 
Consequently, the controller switch ensures that the IDT and energy input trajectories remain strictly orthogonal.

\subsubsection{Heating controller}
\label{sec:heating_controller}
The heating control mode updates the energy input once per layer based on the ILT measurement.
The control algorithm from the study of \cite{kavas2023layer} is adopted, and the laser power is used as the energy input parameter.

The simplified model takes the form: 
\begin{subequations}\label{eq:controller_lti}
	\begin{align}
	\xi_{k+1} &= A \xi_k + B u_k, \\
	y_k &= C \xi_k,
\end{align}
\end{subequations}

where $\xi \in \mathbb{R}^{n_{\xi}}$ is an arbitrary state vector for the model, $u_k \in \mathbb{R}_{+}$ is the single power input applied for the layer, and $y_k$ is the average surface temperature.
The matrices A, B, and C are the identified model matrices obtained from experimental layer-to-layer temperature data~\cite{kavas2023layer}.
Equation \eqref{eq:controller_lti} simplifies most of the complex process dynamics into a linear form.

With the simplified model, the model-based optimization problem is described as:
\begin{align}
	\label{eq:opt_prob_lin}
	\min_{u\in\mathcal{U}}~&  \| A x_k + B u - r\|_Q^2 + \|u\|_R^2,
\end{align}
which is a greedy control objective to minimize the error in the next time step. 
Putting everything together,

\begin{equation}
	\label{eq:l2l_controller}
	u_{k} = \Pi^P_{\mathcal{U}}\left[u_{k-1}  - \alpha P^{-1} \left( \left(CB\right)^TQ ( \hat{y}_{k} -r  ) + R u_{k-1}      ) \right) \right],
\end{equation}

where $P$ is the preconditioner matrix, $\Pi$ is the weighted projection operator to the input (laser power) constraint, \(x_k\) is the process state at layer \(k\) (corresponding to the 
identified model state \(\xi_k\)), \(u\) is the decision variable representing the laser power assigned for the next layer, \(Q\) and \(R\) are positive-definite matrixes that penalize deviations from the reference ILT and changes in the control input to balance aggressiveness and smoothness, respectively, \(r\) is the reference ILT value selected for each geometry based on the pre-study, and $\hat{y}_{k}$ is the estimate considering measurement noise and disturbances.

Further details concerning the preconditioner matrix and projection are described as in~\cite{kavas2023layer}.
To deal with measurement noise and bias introduced by the material due to alternating hatch orientations, a simple moving average filter with window size $w \in \mathbb{N}$ is used:
\begin{align}
	\label{eq:mov_avg_temp}
	\hat{y}_k = \frac{1}{w} \sum_{i = k-w}^k T_{surface}(i),
\end{align}
where $T_{surface}(i)$ denotes the average surface temperature at layer $i$, which is evaluated using the raw measurement signal of the thermal camera over a region of interest as shown in Figure~\ref{fig:recoater_effect}.

The update~\eqref{eq:l2l_controller} forms an approximate gradient and a Hessian-like preconditioner by using the approximate model information and the true model output given by $y$. 
This makes the controller robust to model mismatches. 
The stability of similar controllers under comparable assumptions is given in the context of feedback optimization~\cite{colombino2019online,hauswirth2024optimization} and iterative learning control~\cite{balta2024iterative,spector2018passivity}.
Due to the very thin layer thickness compared to the remaining mass and the cross-sectional area for a layer, even the largest changes in the cross-section do not change the state of the process significantly.
Therefore, controller update is a sufficient approximation to iteratively solve~\eqref{eq:layer_dyn_temp}.

\subsubsection{Cooling controller}

The limit imposed by the melting and solidification phenomena and further microstructural constraints on the laser energy input assignment narrows the actuation range of $\mu_k$ given a positive $\Delta a$, which forces the process to yield defective parts or failure to compensate for the overheating behavior.
This study investigates the effect of $\tau_{\ell}$ and $\Delta a$ in conjunction with the described power controller.

\begin{figure}[ht!]
\begin{center}
\includegraphics[width=0.7\columnwidth]{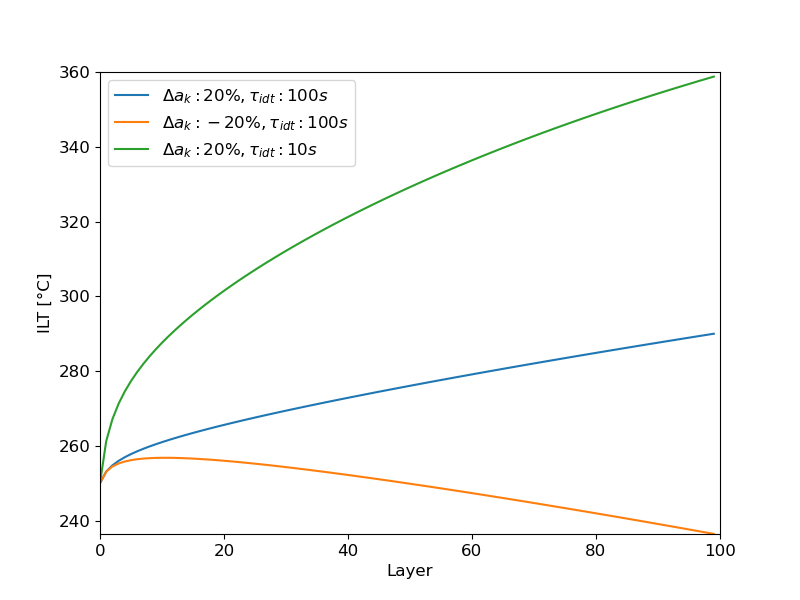}
\caption{Effect of the increasing and decreasing cross-section area and the interlayer dwell time addition, simulating \Cref{eq:layer_dyn_temp} with constant laser power input.}
\label{fig:analytical_simulation}
\end{center}
\end{figure}

Three cases for describing the effect of the IDT and the area rate-of-change are given in \Cref{fig:analytical_simulation} based on \Cref{eq:layer_dyn_temp}. 
For the values of the illustrative parameters shown in the plot legend, negative and positive $\Delta a$ represent converging, upward facing, and diverging, overhanging features, respectively, as percentages.
It is aligned with the intuitive explanation of each case that the ILT exhibits a decreasing trend for the converging case and an increasing trend for the diverging case.
The effect of added waiting time is also observed to decrease the rate at which the diverging cross-sectioned part overheats. 
Although explaining the layer-to-layer temperature change of ILT with the description mentioned earlier, the within-layer cooling dynamics are critical for the $\tau_{IDT}$ assignment.

In a standard PBF-LB/M process, no active cooling is available, and the part geometry, orientation, and support structures are fixed. 
Cooling therefore relies solely on passive conduction and convection, with each layer providing a constant recoating time $\tau_c(k)$ that also acts as a cooling period. 
However, constant cooling is insufficient: simulations based on \Cref{eq:layer_dyn_temp} and prior work \cite{kavas2023layer} show that overheating still occurs. 
To address this, an additional dwell time $\tau_{idt}$ is introduced and triggered from in-situ thermal images of the top surface. 
The algorithm pauses the next layer until the ILT drops to the reference value, ensuring that only the minimum necessary $\tau_{idt}$ is added under the considered control architecture.

%% file: 3-Material_and_Methods.tex
\section{Materials and Methods}
\label{sec:materials_methods}

In this section, the experimental setup used to implement the proposed controller are described. 
These include the PBF-LB/M machine, and the corresponding experimental parameters, specimen design, controller implementation, and methods of specimen characterization.

\subsection{Experiment design}
\label{sec:experiment_design}

Four sets of experiments were conducted to evaluate the proposed method. 

\begin{itemize}
    \item A geometry featuring diverging and converging cross-sectional area regions was built to evaluate the performance of the switched control architecture in both overhanging and upward-facing scenarios, where net heat accumulation and net dissipation are anticipated, respectively. 
    \item The diverging and converging geometry was printed three more times in two separate build jobs without the switched control. 
    The first build had one part that utilized solely laser power control, together with a second part as an uncontrolled reference. 
    The second build had one uncontrolled part printed with a static IDT assignment of 15 seconds.
    \item A breakaway line support structure was incorporated into the same inverted pyramid design to assess its effect on mitigating the overheating due to the cross-section change rate. 
    \item The switched control algorithm was applied to an inverted pyramid design with a 45° overhanging angle.

\end{itemize}

\subsubsection{Specimen Design for Heat Accumulation}
\label{sec:specimen_design}

\begin{figure}[ht!]
\begin{center}
\includegraphics[width=0.7\columnwidth]{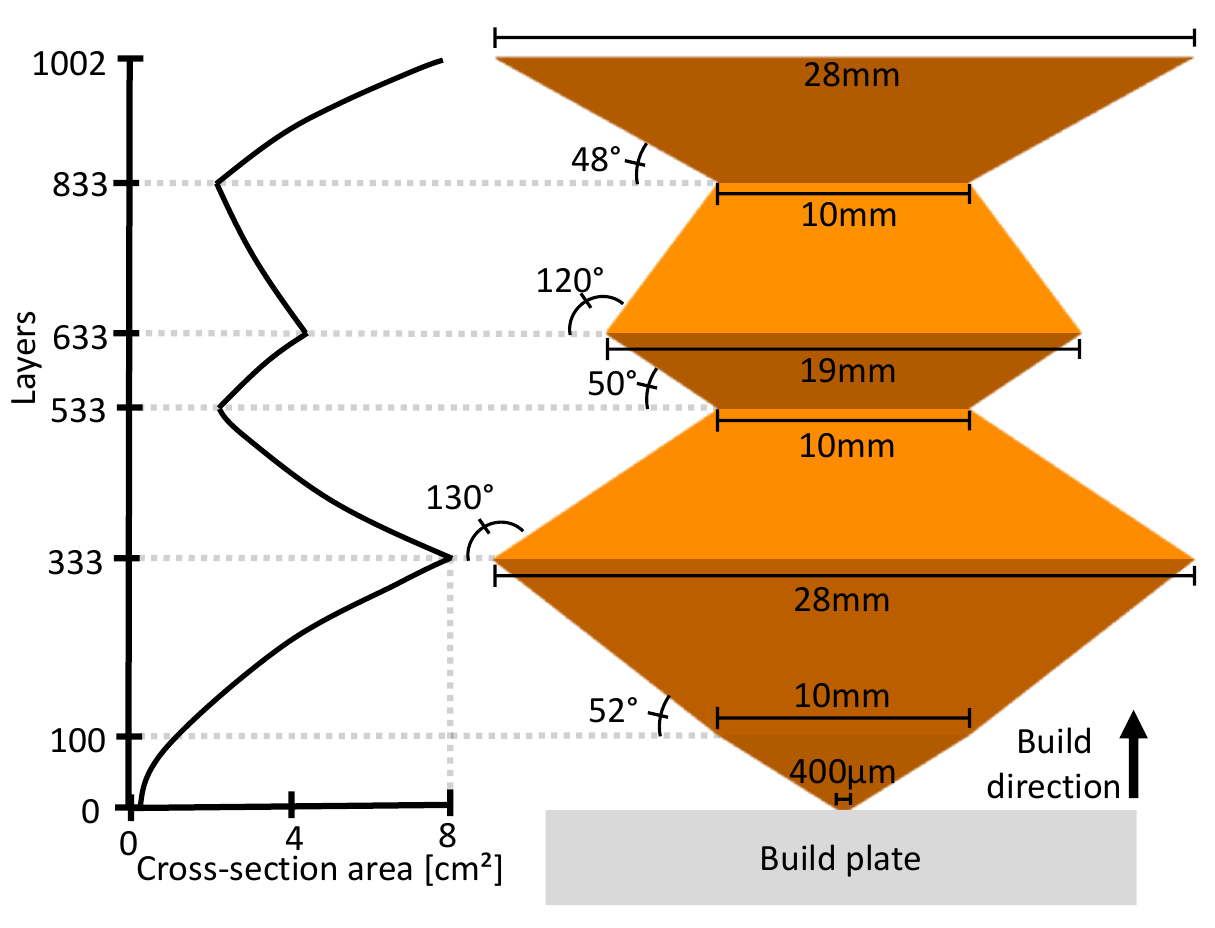}
\caption{The specimen design with converging and diverging cross-sections.
The cross-sectional area progression was quantified by the plot on the left with corresponding dashed lines for each change rate transition.
Up and downfacing angles were shown in the image on the right.
}
\label{fig:totem_part_design}
\end{center}
\end{figure}

Figure \ref{fig:totem_part_design} shows the diverging and converging cross-section part design.

The designed part consists of three diverging sections and two converging sections. 
The overhanging angles of the diverging sections were 52°, 50°, and 48°, while the angles in the converging sections measured 130° and 120°, respectively. 
Although the exact angles were chosen arbitrarily, the smallest section for precise thermal camera readings was assumed to be 10 mm². 
The part had a total of 1002 layers and a height of 30 mm.
Although the build plate is operated in a heated configuration, the connection to the platform is intentionally minimized: the bottom connection of the part has a 400~\textmu m $\times$ 400~\textmu m square cross section to promote a quasi-steady thermal boundary condition, i.e., to limit conductive heat loss to the build platform.

The trends in cross-sectional area changes are presented in the plot on the left side of the figure, which corresponds with the results. 
The main difference between this part and the single inverted pyramid lies in the first diverging section, where the connection to the build plate was smaller to facilitate removal and enhance heat accumulation. 
This connection node measured 0.4 mm²; due to its smaller size, the 10 mm² minimum surface area was achieved within 100 layers while avoiding a very low overhanging angle. 
Consequently, the controller initialization layer was set to 100 for the converging-diverging cross-section part.

\begin{figure}[ht!]
\begin{center}
\includegraphics[width=1\columnwidth]{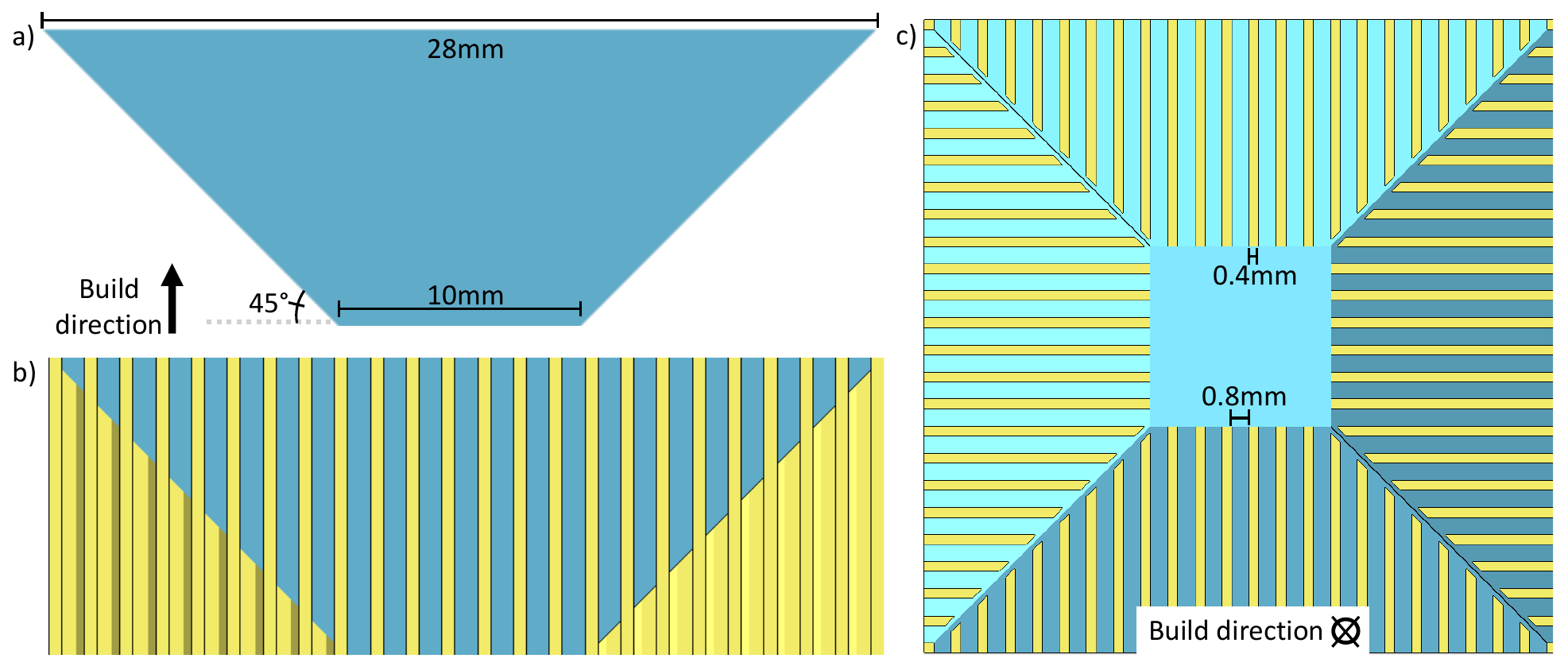}
\caption{(a) The side view without the support structure, (b) side view with the support structure, and (c) the bottom view of the support design.
The breakaway supports were designed with a thickness of 0.4mm and separated by 0.8mm.
No taper was added to the connection with the part.
Line supports were not connected at the edges to maintain brake-away capability.
}
\label{fig:inverted_pyramid_part_designs}
\end{center}
\end{figure}

The inverted pyramid and the support structure design are shown in \Cref{fig:inverted_pyramid_part_designs}.

\Cref{fig:inverted_pyramid_part_designs} a) illustrates the inverted pyramid without any supports; b) presents the added support structures from the same perspective; and c) depicts the supports viewed from the bottom to the top, as indicated by the build direction legend. 
The line supports were designed with a uniform thickness of 0.4 mm and extruded until they merged with the pyramid's surface. 
The line-to-line separation was set to 0.8 mm, and the lines were intentionally not connected at the edges of faces with different orientations to facilitate their removal as breakaway supports.
The support structures are processed with a static laser power throughout the build, even though the laser power assigned to the part itself is actively controlled.
The parts were 10 mm in height, resulting in a total of 333 layers after slicing.

The parts were designed using NX 8 (Siemens) and exported in STL file format. 
They were then imported into Autodesk Netfabb (Autodesk Inc.) and positioned on the build platform within the CAD environment. 
The geometries were trapezoidal prisms, symmetrical in their XZ and YZ planes. 
The parts were printed symmetrically relative to the thermal camera's field of view at the center of the build plate.

\subsection{ILT \& IDT Estimation}
\label{sec:estimation}
The proposed parameter fitting approaches in \Cref{sec:heat_buildup} are applied by using the first 200 layers of ILT data (layers between 100 to 300) of the 15s static IDT assigned part. 
The identified model with constant $\hat{\mathcal{K}}_k$ and layer-dependent $\hat{\mathcal{K}}_k(A_i)$ values are then used to estimate the ILT throughout the remainder of the part. 
The fitting layer range is deliberately restricted to a single overhanging region without transitions (see \Cref{fig:totem_part_design}), in order to avoid overfitting to a specific geometry used for model identification. The element height $\Delta l$ is selected manually to ensure a stable fit of $\hat{\mathcal{K}}_k(A_i)$ and physically plausible ILT predictions.
Furthermore, the identified model is applied to estimate the IDT profile required for each layer to cool to the selected reference ILT, and the resulting profile is compared with the IDT values of the switch-controlled part.

\subsection{Laser Powder Bed Fusion Test Bed}

The experiments were conducted using an Aconity3D Midi+ (Aconity3D GmbH) PBF-LB/M machine \cite{aconity3d}.
A $20~mm$-thick stainless steel 1.4301 build plate was used for all experiments with the heated build platform temperature set to 200$^{\circ}$C. 
The layer height was set to $\SI{30}{\micro\meter}$ and a silicone recoater blade was used. The processing laser was a 1080 nm continuous wave Gaussian mode fiber laser with up to 500W output (nLIGHT Alta), which was focused down to a beam diameter of $\SI{80}{\micro\meter}$ for all builds. The powder used in this study was gas-atomized stainless steel 316L (1.4404) with a $d_{10}$-$d_{90}$ particle size distribution of $15-\SI{45}{\micro\meter}$ (CT POWDERRANGE 316LF, Carpenter Additive).

\subsection{Parameter window selection}
\label{sec:parameter_window_selection}
The nominal parameters for laser power, scan velocity, and layer thickness are presented in Table \ref{parameters_table}, detailing both hatch and contour scans, as well as hatch properties. 
For the variable laser power approach, it is essential to identify the lower and upper threshold values. 
Numerous studies have indicated that the lower threshold for energy input per volume corresponds to the onset of lack-of-fusion porosity, while the higher threshold marks the beginning of the keyhole melt pool mode \cite{lee2016simulation}.
Here, the lower and the upper bound were set to $\SI{140}{\watt}$ and $\SI{200}{\watt}$, respectively.
Based on the switching algorithm, the lower bound for the laser power ($\SI{140}{\watt}$) is defined as $P_{low}$, the value at which the algorithm initiates cooling once it is reached.
It was experimentally validated that a microstructure without significant lack-of-fusion pores can be achieved at the lower boundary and without keyholing at the upper boundary of laser power. 
The hatch orientation for the linear scan vectors started at 0 degrees (aligned with the X-axis) and was rotated by 90 degrees in each subsequent layer.

\begin{table}[htbp]
  \centering
  \caption{Parameters used for the processing of SS316L steel}
  \label{parameters_table}
  \begin{tabular}{@{}*{7}{p{2cm}}@{}}
    \toprule
    Parameters & Nominal Laser Power [W] & Scanning Speed [mm/s] & Hatch Spacing [$\mu$m] & Layer Thickness [$\mu$m] & Initial Hatch Orientation[$\degree$] & Hatch Rotation [$\degree$] \\
    \cmidrule(lr){1-7} \\[-1em]
    Hatch & 150 & 800 & 100 & 30 & 0 & 90 \\
    \bottomrule
  \end{tabular}
\end{table}

The reference surface temperature value was chosen to achieve the widest dynamic range of power input before saturating to either the upper or lower input power thresholds, which was determined experimentally during a pre-study. 
The reference temperatures of $\SI{620}{\celsius}$ and $\SI{650}{\celsius}$ were set respectively for the supported-unsupported inverted pyramid parts and the diverging-converging parts.
For the diverging-converging part, a higher reference temperature was assigned because of the observed elevated and more rapidly increasing ILT.
Further discussion on the criteria for reference temperature assignment was shared in Section \ref{sec:results}.\\

\subsection{Controller implementation}
\label{sec:implementation}
\begin{figure}[ht!]
\begin{center}
\includegraphics[width=0.8\columnwidth]{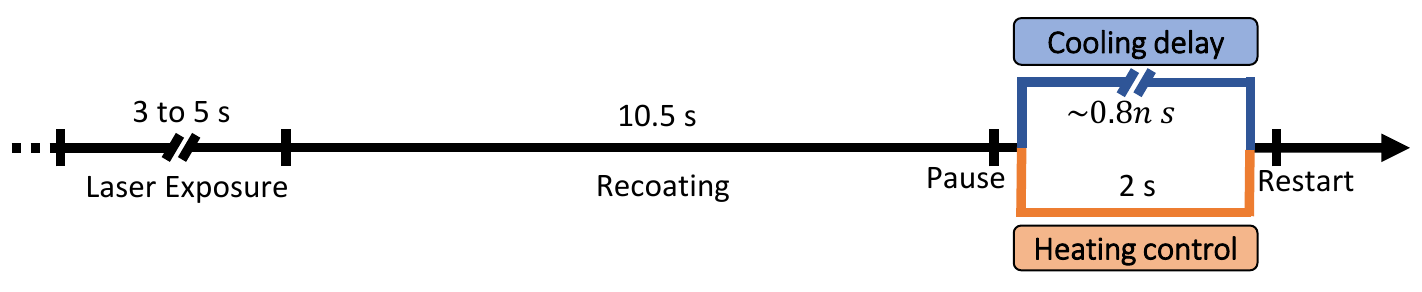}
\caption{Process step timeline within a single layer.
Laser exposure time depends on each layer's cross-section area. The machine recoat time was 10.5 seconds, after which the process was paused for controller processing and execution. If the heating mode is active, it takes approximately two seconds to restart the build and with the cooling mode active, each the camera snapshot cycle takes 0.8 seconds.}
\label{fig:Process_time_chart}
\end{center}
\end{figure}

The controller, data processing, and build job scheduling algorithms are developed in-house and are run on a Python 3.7 environment that is in the main machine control PC.
The provided Application Programming Interface (API) by the machine manufacturer is used to interact with the build job, process the sensor data, and update the necessary parameters such as laser power and IDT.
A detailed flow diagram of a single-layer application is shown in Figure \ref{fig:Process_time_chart}. 

The main timeline for a full layer consists of three phases: laser exposure, recoating, and controller execution. 
In this study, laser exposure for the selected parts took between 3 to 5 seconds, depending on the changing exposure area throughout the layers. 
After laser exposure, the recoating process was initiated, lasting 10.5 seconds. 
Once recoating was complete, the build job paused, and the algorithm executed the control step, depending on whether the heating or cooling mode was required.
The temperature measurements are performed on single-frame snapshots in both heating and cooling cycles, without any serial video recording.

The heating control phase took approximately 2 seconds, with the majority of time spent on initialization and communication to update the parameters, while the controller's calculation time was negligible. 
After updating the laser power input, the next layer began with the laser exposure phase. 
The model used in the laser power controller implemented in this study was the same model identified in \cite{kavas2023layer} in the state-space form defined in (\ref{eq:controller_lti}) on the experimental data of a simple diverging structure with a static overhanging rate.

During the cooling mode, the algorithm waits until the reference temperature is reached.
Each decision made by the algorithm takes approximately 0.8 seconds, including the thermal camera snapshot, data read and transfer, and temperature calculation.
Therefore, the total time spent was a factor $n$ of 0.8 seconds for the cooling mode.

\subsection{Measurements and Sensor Calibration}
\label{sec:feedback_measurements}

The thermal camera used in this study was an Infratec HD800 positioned above to the side of the build plate, looking down with an inclination of $\SI{74}{\degree}$ to the horizontal plane. 
The distance between the center of the platform to the center of the camera lens was $\SI{42}{\centi\meter}$. 
The thermal camera was capable of full-frame image capture with three different aperture settings. 
For this study, the aperture settings were set for the range $\SI{0}{\celsius}$ to $\SI{500}{\celsius}$ per the manufacturer's recommendation.
The camera has a measurement accuracy of $\pm\,1.5~\mathrm{K}$.
Although the aperture setting is recommended for measuring objects with up to $\SI{500}{\celsius}$, based on all the singal responses recorded, no saturation was observed as the time period of interest is relatively long after the laser scanning.
For a fast control decision cycle, the measurement performance was critical in the proposed control architecture.
Rather than full-frame acquisition of the camera, a quarter-frame was used to reduce the snapshot size from 1024 × 768 pixels to 512 × 384 pixels, which also reduced the time to acquire and process the images. This has no influence on the pixel size or feature resolution of the image.
The exact ILT calculation approach employed and the further setup details are the same as described in our previous study~\cite{kavas2023layer} and are not repeated here for brevity.

\begin{figure}[ht!]
\begin{center}
\includegraphics[width=0.8\columnwidth]{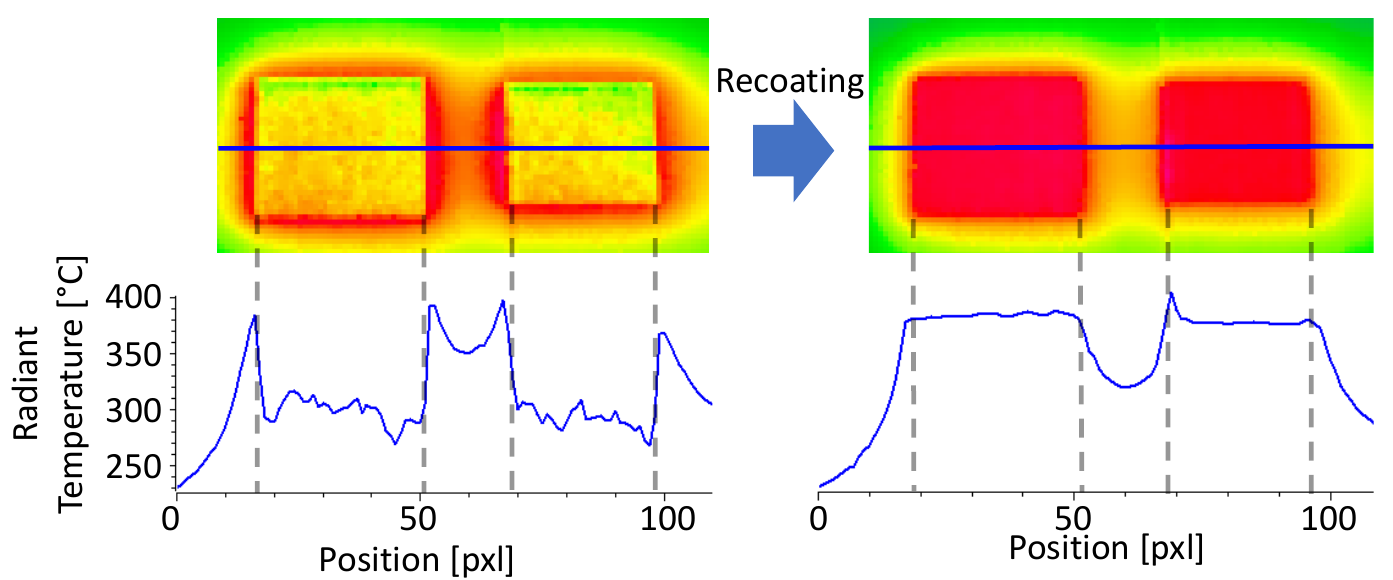}
\caption{Effect of powder on the radiant temperature acquisition: pre- and post-recoating emission difference.}
\label{fig:recoater_effect}
\end{center}
\end{figure}

The critical detail employed in this study regarding the temperature acquisition is measurement through the recoated surface. 
During a layer cycle, the first snapshot from the thermal camera was taken after the recoating period, with a layer of powder applied over the surface of the part rather than from the exposed metal surface. 
This design choice was made for two main reasons.
First, the definition of ILT requires access to the temperature of the part's surface constant time before the exposure of the next layer, and this temperature is most accurately measured immediately before exposure. 
Since this study focuses on the addition of IDT, a more accurate and layer-to-layer comparable ILT reading is achieved after recoating.
Second, there is a strong dependence on the hatching orientation, which results in higher variance in temperature readings from each pixelated location on the part's surface. 
A thin layer of powder was found to diffuse the emissions from the part beneath the powder surface while also homogenizing the radiant temperature captured in the thermal image, as illustrated in Figure~\ref{fig:recoater_effect}.

On the left of the image, a thermal image was captured after exposure of two parts with square cross-sections.
The right side of the image shows the same parts, immediately after recoating for the next layer.
It was observed that the powder-coated surfaces show a much more homogeneous profile.
A constant layer thickness of powder has less noise since the surface condition-related perturbations, such as surface color, vector orientation, and roughness, were eliminated~\cite{lane2015calibration}.
Furthermore, the temperature signal is observed to be higher on the recoated surface.
A similar effect is also observed in the periphery of the solid areas in powder in the left (not recoated) image.
This stems from an anecdotally well-known (but less documented) phenomenon stemming from the high reflectivity and low emissivity of the relatively smooth and shiny top surface of the part.
Given the incidence angle of the camera relative to the flat uncoated surface, the signal is strongly biased by reflections, resulting in an inaccurate measurement.
The thin powder layer acts to prevent reflections and, due to the sperical shape of the powder, tends to radiate equally in all directions, resulting in a more accurate  and consistent measurement.
The powder on top of and immediately surrounding the part is heated to approximately the same temperature as the part itself, which is apparent from comparing the images of \Cref{fig:recoater_effect} without and with powder.

\begin{figure}[ht!]
\begin{center}
\includegraphics[width=0.8\columnwidth]{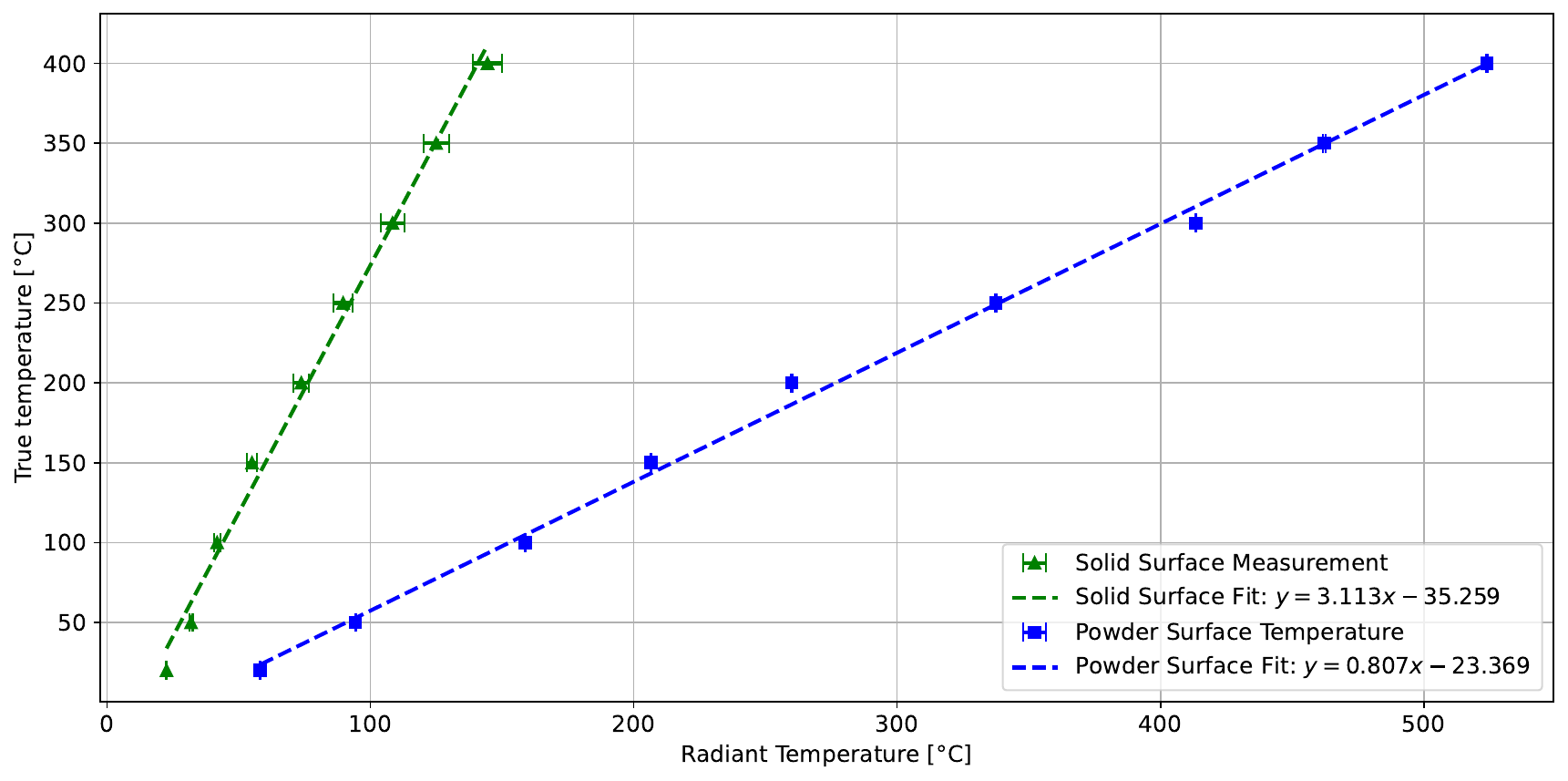}
\caption{Calibration of thermal camera signals using a controlled heating setup up to 400 °C. The resulting calibration curve for the powder-recoated surface is compared with that of the solid part surface.}
\label{fig:calibration_curve}
\end{center}
\end{figure}

The camera was calibrated using the same setup documented in our previous study \cite{kavas2023layer}, this time also with a thin layer of powder on top of the cubes. 
The signals acquired in the thermal camera data were calibrated by heating the build plate its resistance heater assembly (provided by the machine manufacturer) from room temperature to 400°C in steps of 50°C.
The system was held for five minutes at each measurement point to allow sufficient time for heat to distribute and stabilize across the build plate.
The surface temperature of the cubes was measured using thermocouples inserted in holes $2~mm$ below the top surface and seated with conductive paste.
\Cref{fig:calibration_curve} shows the resulting calibration curves taken without (green) and with (blue) the thin layer of powder.

The X-axis, as radiant temperature, describes the raw temperature values measured by the camera, with no emissivity gain applied~($\epsilon=1$).
Therefore, the emissivity source is the black body radiation of the printed part that is elevated to the described controlled temperature.
The Y-axis, as the true temperature, describes the actual temperature value that the build plate heating system is set to.
The functions that map the radiant to true temperature are calculated by least squares with a linear fit and described in the legend as solid and powder surface fit equations for calibrating the measured temperature values to the actual value of the solid surface temperature.
The difference in variance, as indicated by the error bars at each calibration temperature, shows that the powder-recoated surface provides less noisy data compared to the solid surface.
The obtained fit equations are used to calibrate the raw measurements for all the experiments described.

\subsection{Characterization of the specimens}
\label{sec:characterization}
Specimens were removed from the build plate by electrode-discharge machining. 
They were cut close to their center line by a Struers Accutom-10 cutting machine, aligned to the movement of the recoater, then metallographically prepared by embedding, grinding, and polishing.
Embedding was performed hot by a Struers CitoPress machine with DuroLite bakelite resin grinding with 320-grit sandpaper, and polishing with Struers commercial polishing cloths Largo, Dac, Nap, and Chem with suspensions with particle sizes of $\SI{9}{\micro\meter}$, $\SI{3}{\micro\meter}$, $\SI{1}{\micro\meter}$, and $\SI{0.1}{\micro\meter}$, respectively.
Polished mounts were scanned by a Keyence VHX-7000 microscope in coaxial and ring lighting modes to observe the microstructure.

%% file: 4-results_discussion.tex
\section{Results}
\label{sec:results}

\subsection{Converging \& diverging cross-sections}
\label{sec:Results_converging_diverging}
\begin{figure}[ht!]
\begin{center}
\includegraphics[width=1\columnwidth]{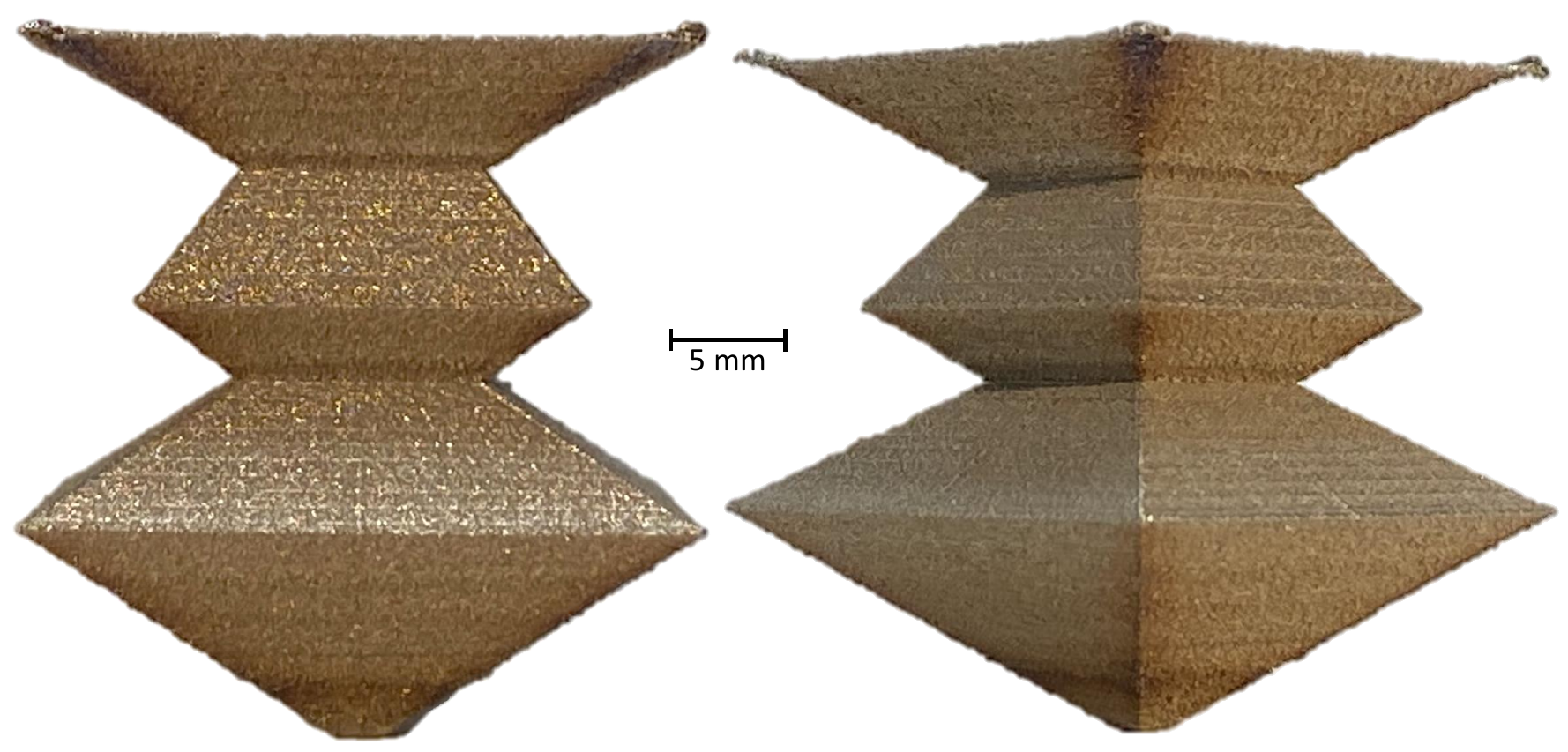}
\caption{The printed part profile images from the side (on the left) and from the diagonal orientation (on the right) of the diverging and converging cross-sectioned parts with the switched heating/cooling controller active.}
\label{fig:totems_printed}
\end{center}
\end{figure}

\begin{figure}[ht!]
\begin{center}
\includegraphics[width=1\columnwidth]{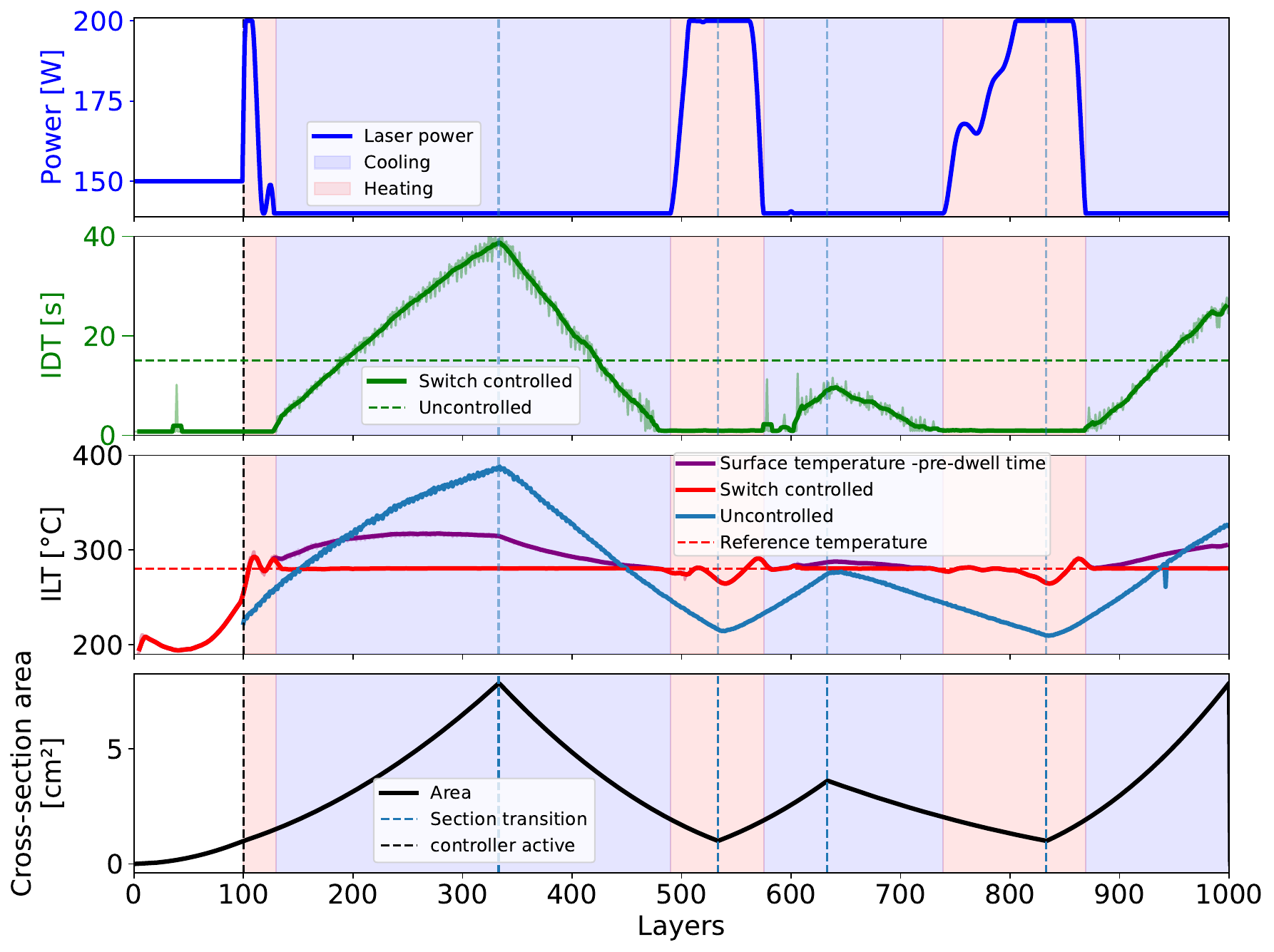}
\caption{ILT, IDT, and power assignment data plotted with the cross-sectional area progression. Cooling and heating control mode layer regions are highlighted in blue and red, respectively and the diverging-converging transitions are shown with the vertical dashed blue lines.}
\label{fig:totem_plot}
\end{center}
\end{figure}

Pictures of the printed diverging and converging cross-sectioned parts are given in Figure~\ref{fig:totems_printed}.
The left image is the side view and the image on the right is the diagonal view of the specimen produced using the switched heating and cooling controller, with the build direction upward. 
The ILT, laser power, and IDT assignments corresponding to this part are given in Figure~\ref{fig:totem_plot}.

In the first subplot, laser power is given and the second subplot is comprised of the IDT assignments.
The third subplot shows the ILT measurements.
The purple line shows the temperature measurement before applying IDT while the red line shows the ILT measurement immediately before the exposure of the next layer for the controlled part.
Naturally, the purple line corresponds to the ILT; therefore, the red and purple lines merge where there is no IDT applied, or where the heating mode is active.
The fourth subplot shows the cross-section progression of the part.
The vertical dashed blue lines show the layers at which the cross section transitions from converging to diverging and vice versa.
All plots are also highlighted based on the active control mode.
Blue regions represent layers where the cooling controller was active and an IDT was applied.
Red regions show where the heating controller was active and the laser power was varied.

\begin{figure}[ht!]
\begin{center}
\includegraphics[width=1\columnwidth]{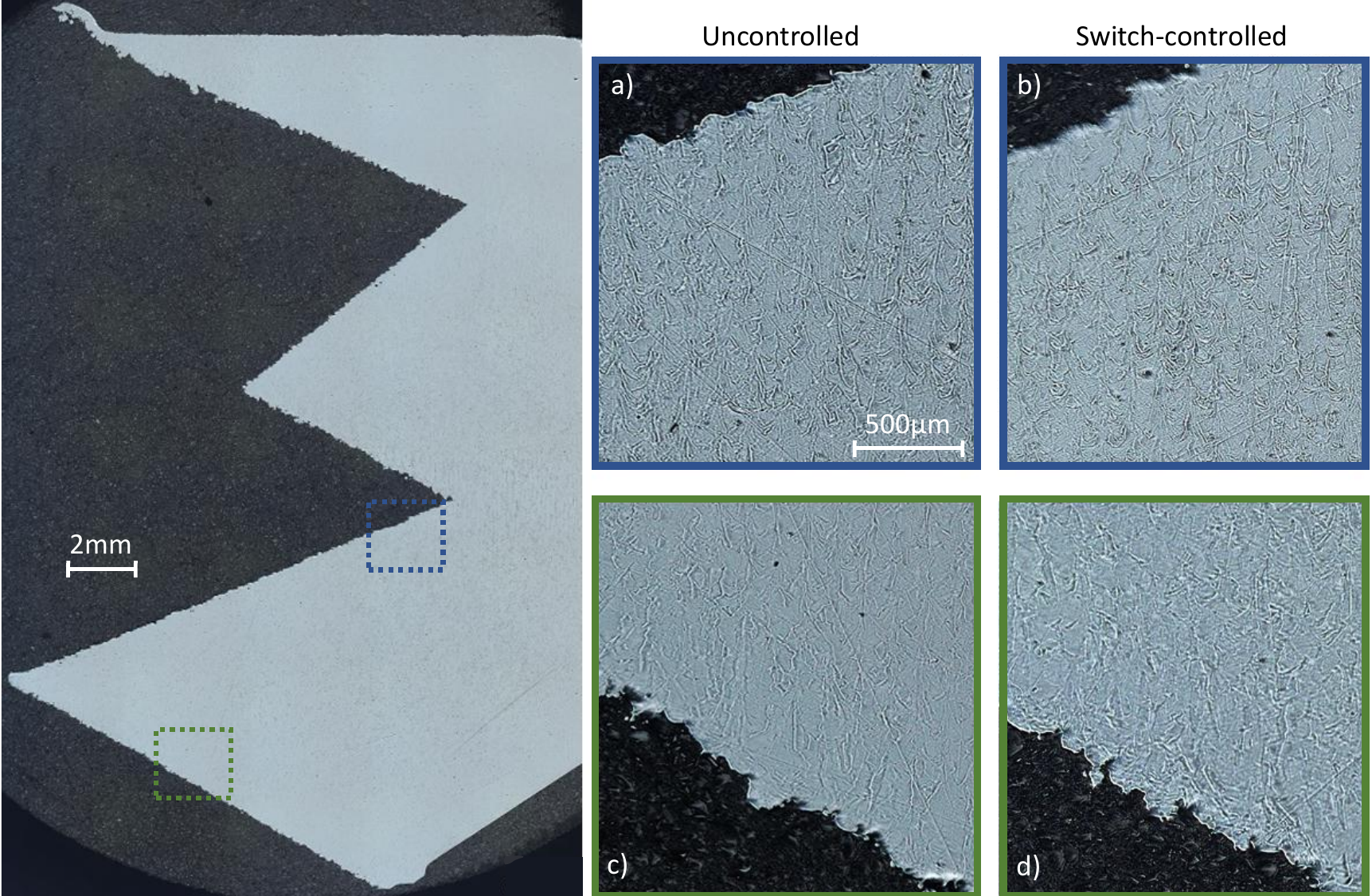}
\caption{Cross-sectional micrographs of the switch-controlled and uncontrolled parts. 
The left image shows the cross-section of the switch-controlled part; magnified views (b) and (d) are extracted from this cross-section. 
Magnified views (a) and (c) correspond to the uncontrolled part. 
The uncontrolled build could not be printed to the same number of layers; therefore, magnified images from higher layers are not available for comparison.}
\label{fig:microstructure}
\end{center}
\end{figure}

Microstructure images of the metallographic investigation of the switch-controlled and uncontrolled parts are shown in \Cref{fig:microstructure}.

The image on the left shows the cross-section of the switch-controlled part; the magnified views in (b) and (d) are taken directly from this cross-section.
The magnified views in (a) and (c) correspond to the uncontrolled print.
Since the uncontrolled build could not be printed for the same number of layers, a comparison at higher layers was not possible.
Parts are cut on the diagonal axis, hence the significant corner tip bending due to the overhanging angle that is visible in \Cref{fig:totems_printed} is also observed in the upper left-hand side region of \Cref{fig:microstructure}.

Qualitative observations indicate that, near the cross-sectional change, the uncontrolled region in (a) exhibits relatively consistent grain sizes, whereas the controlled region in (b) shows locally coarser grains in the upper-right area, where higher laser power was applied by the controller.
In contrast, the uncontrolled sample exhibits elongated grains closer to the overhang, while the controlled sample shows a more uniform grain morphology in the corresponding region.

\begin{figure}[ht!]
\begin{center}
\includegraphics[width=1\columnwidth]{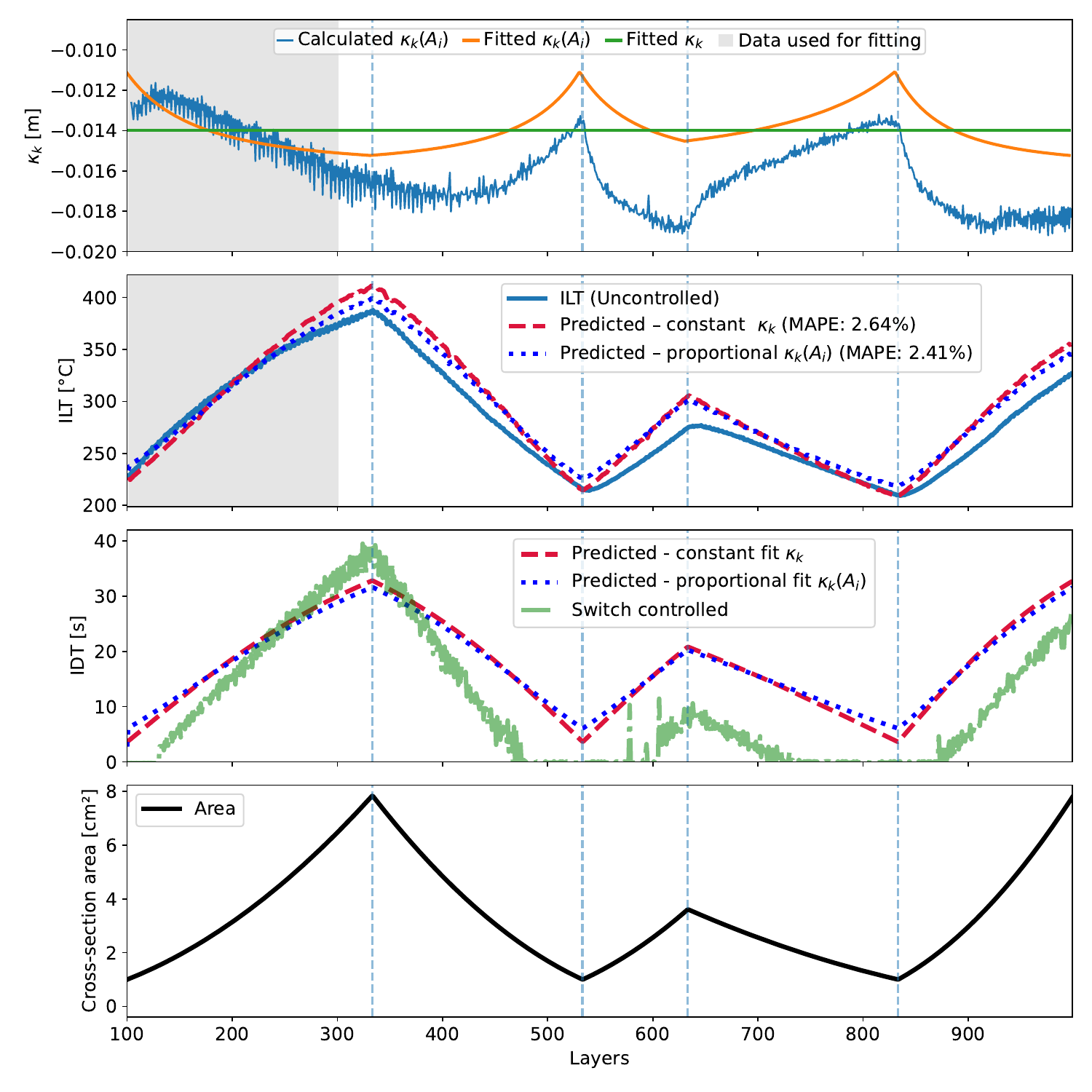}
\caption{Fitted constant $\mathcal{K}_k$, layer-dependent  $\mathcal{K}_k(A_i)$ , and ground-truth $\hat{\mathcal{K}_k}$ (row 1); predicted ILT vs.\ measurements of the 15,s static reference part (row 2); IDT predictions (row 3); and area progression per layer (row 4). Grey regions mark the fitting range, and vertical dashed lines indicate diverging/converging transitions.}
\label{fig:idt_prediction}
\end{center}
\end{figure}

The modelling approach described in \Cref{sec:heat_buildup} is applied per \Cref{sec:estimation} as given in \Cref{fig:idt_prediction}.

The first subplot shows the fitted constant $\mathcal{K}_k$ and layer-dependent $\mathcal{K}_k(A_i)$ values, as well as the calculated actual $\hat{\mathcal{K}_k}$ values as the ground truth for each layer. 
The grey highlighted region in the first 200 layers of the plots shows the data range that is used for model fitting.
The second row shows the predicted ILT values with both parameter fitting approaches, as well as the actual measurements of the 15~s static IDT assigned reference part.
The third row shows the IDT predictions based on both models. 
The last row, same as the \Cref{fig:totem_plot}, shows the area progression per layer.
The vertical dashed lines on each plot show the diverging \& converging transitions based on the last row. The ILT prediction accuracy of the constant  $\hat{\mathcal{K}_k}$ has been quantified as a MAPE 2.64\%, and of the proportional  $\mathcal{K}_k(A_i)$ as a MAPE 2.41\%. The total predicted IDT is 6 hours and 15 minutes while the actual IDT is 4 hours and 48 minutes.

\subsubsection{Reference build for converging \& diverging cross-section}
\label{sec:ref_builds}

For comparison, the converging and diverging geometries were produced uncontrolled (static laser power of $150~\mathrm{W}$) and with heating-only control (layer-wise power adaptation), without additional inter-layer dwell time (IDT).
Both builds terminated prematurely due to recoater contact and relevant build data is only obtained until layer 400.
While recoater contact was identified as the failure mode, a definitive causal link to thermal distortion cannot be established.
In a subsequent build, the uncontrolled part was successfully produced by introducing a static $15~\mathrm{s}$ IDT.

\begin{figure}[ht!]
\begin{center}
\includegraphics[width=0.9\columnwidth]{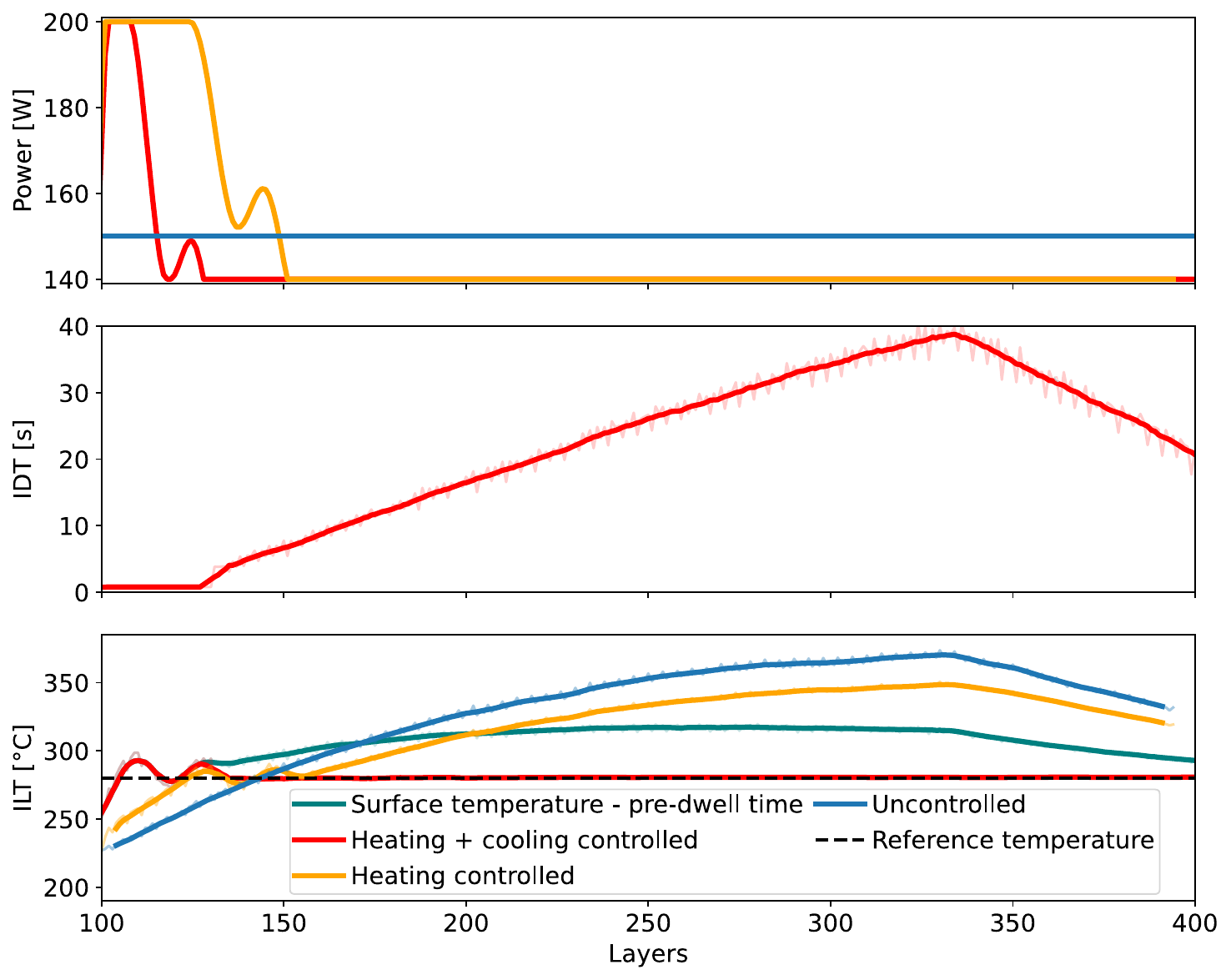}
\caption{
Comparison of the ILT trends and the actuation of the heating \& cooling controlled part, only the heating controlled part, and the uncontrolled part.
The first row shows the power assignment, second row IDT assignment and the third row shows the ILT measurements, including the pre-dwell time surface temperature.}
\label{fig:comparison_plot}
\end{center}
\end{figure}

Due to the early failure of the reference parts, the ILT and actuation comparison is possible only until layer 400.
~\Cref{fig:comparison_plot} compares the evolution of the ILT and the corresponding controller actions for three cases: the combined heating and cooling controller, the heating-only controller, and the uncontrolled process. 
The first row illustrates the layer-wise laser power assignments, while the second row presents the IDT adjustments applied by the cooling controller. 
The third row shows the resulting ILT measurements, including the surface temperature recorded before the IDT period for the heating and cooling-controlled part.
The shallower slope of ILT of the heating-only controlled case up to layer 150 is attributed to the increased effective IDT caused by the simultaneous printing of the additional reference part, whereas the switch-controlled case was printed as a single part with only recoating-induced waiting time prior to the cooling switch.

\begin{figure}[ht!]
\begin{center}
\includegraphics[width=1\columnwidth]{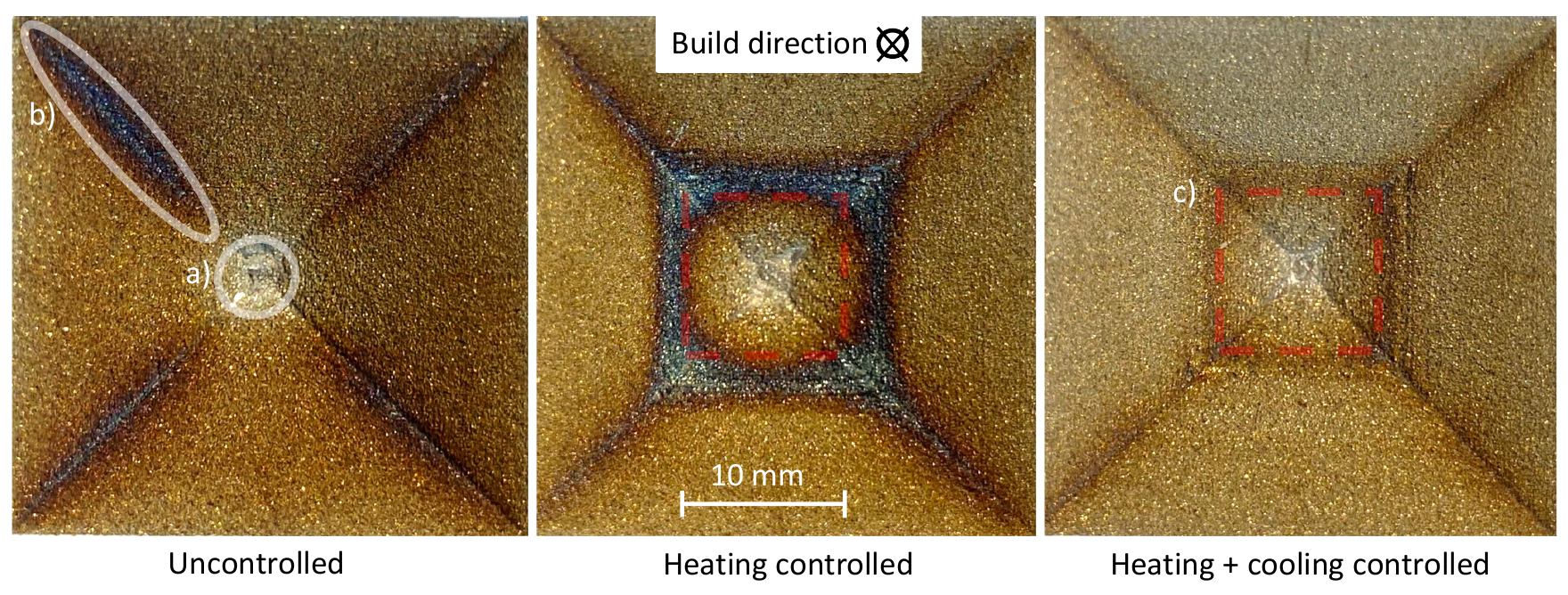}
\caption{The bottom surface images of the switch-controlled (right) and the uncontrolled (left) parts.
a) the build plate connection, b) the edges showing corrosion related to overheating, and c) indicates layer 100 where the controller was initialized. 
}
\label{fig:totem_buttom}
\end{center}
\end{figure}

\Cref{fig:totem_buttom} shows a comparison of the bottom surfaces switch-controlled and uncontrolled diverging and converging parts. Region a) describes the connection with the build platform for all parts, where the remnant of the break-away connection is visible.
Region b) indicates the edge locations with visible discoloration (increased oxidation) due to the overheating behavior observed in the uncontrolled part.
Perimeter c) marks the cross-section of the layer that the controller is initialized at layer 100.

\subsection{Supported versus unsupported overhangs}
\label{sec:results_supported}

\begin{figure}[ht!]
\begin{center}
\includegraphics[width=0.6\columnwidth]{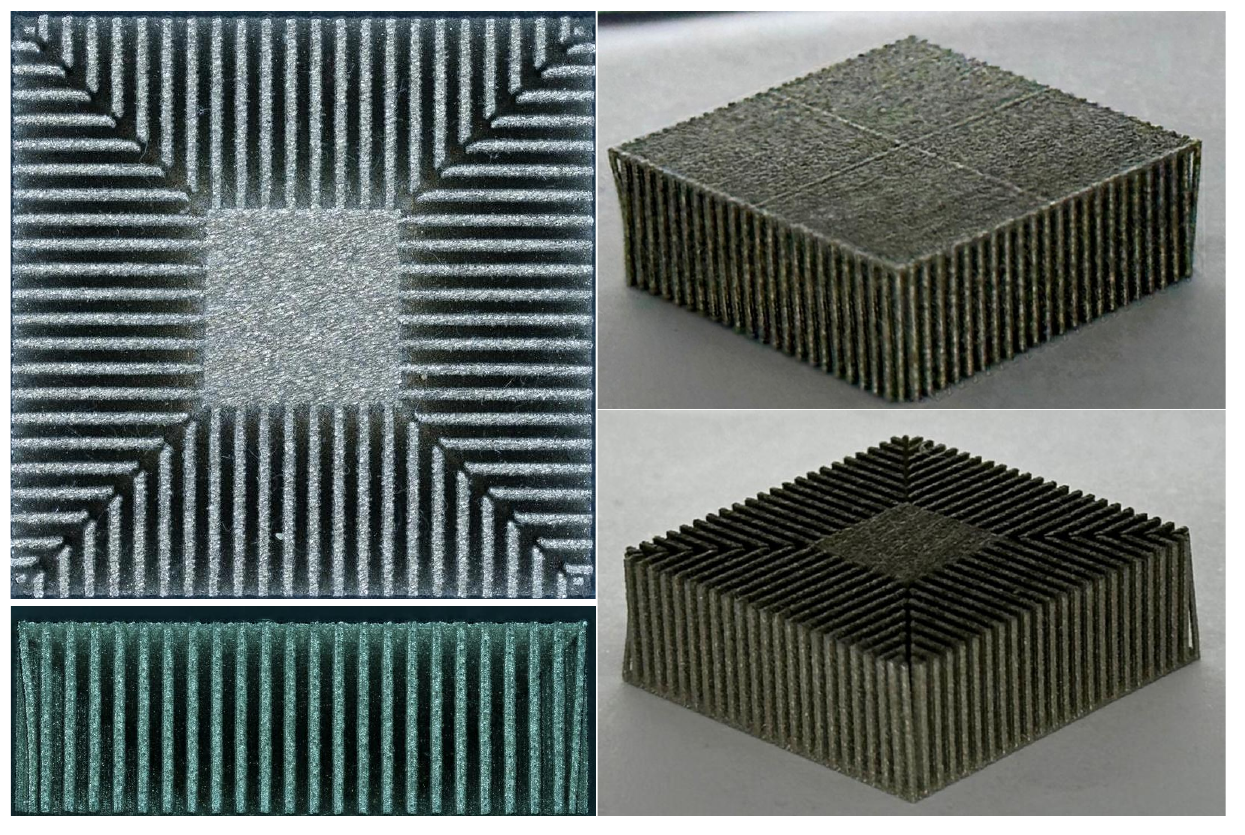}
\caption{Pictures of the printed inverted pyramid part with line supports from bottom (left-hand side up), side (left-hand side down), isometric from top, and bottom directions (right-hand side up and down).}
\label{fig:supported_part_printed}
\end{center}
\end{figure}

\begin{figure}[ht!]
\begin{center}
\includegraphics[width=1\columnwidth]{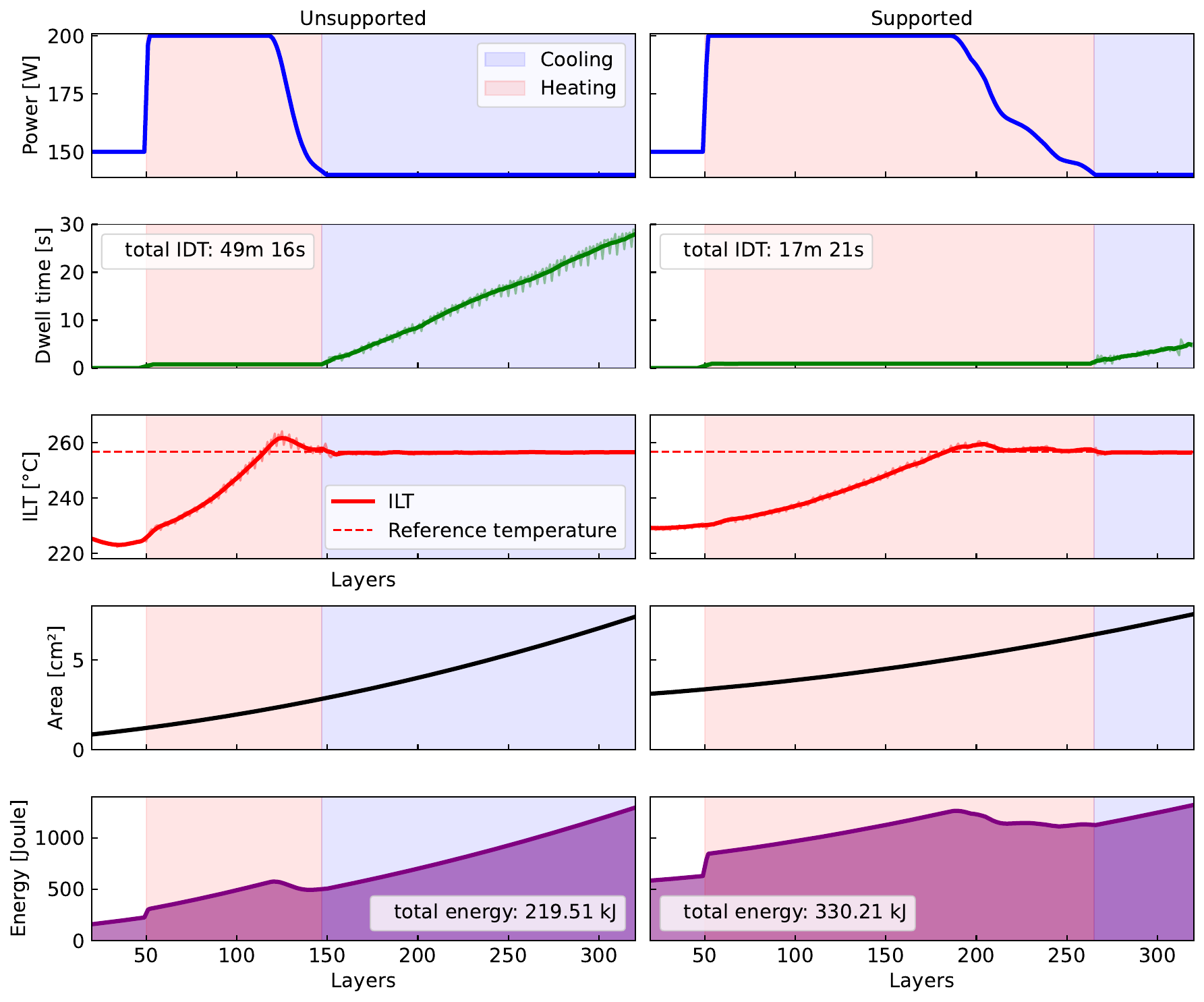}
\caption{The first row shows the laser power assignment, followed by IDT and ILT measurements in the second and third rows, with raw data in light red and green. Rolling averages (window size 8) are in darker shades. The fourth row presents the cross-sectional area per layer, while the fifth row plots the calculated energy consumption ($E$) per layer. Active controller modes, highlighted in light blue (cooling) and light red (heating), are shown across all subplots.}
\label{fig:supported_unsupported_comparison}
\end{center}
\end{figure}

\Cref{fig:supported_part_printed} shows images of the supported part while the corresponding ILT measurements, IDT assignment, laser power assignment, Layer-wise cross-section area, and the energy consumption per layer for the supported and unsupported inverted pyramids are given in Figure~\ref{fig:supported_unsupported_comparison}.

The first row displays the laser power assignment, while the second row contains the IDT data and the third row presents the ILT measurement. 
The raw data for ILT and IDT are highlighted in light red and light green, respectively. 
Total IDT assignments per part are given in the relative plots, calculated by the area under the curves.
Their rolling averages, calculated with a window size of 8, are shown in a denser shade of the same colors.
Forth row shows the cross-section area per layer.
In the fifth row, the calculated energy consumption $E$ is plotted per layer based on 

\[
E = \frac{P}{v} \times L_{\text{total}}
\]

where the $P$ is laser power, $v$ is the scan speed, and $L_{total}$ is the total vector length.
The total energy consumption per part is also given in the plot, desribed by the purple highlighted area under the curves.
The active controller modes as heating and cooling for respective layers are highligted in light red and light blue in all the subplots.

In ~\Cref{mean_increase_table}, several quality criteria and statistical values including the error, rise time, number of layers in heating and cooling modes, and total IDT duration are shown for the supported and unsupported parts.

The 1\% error threshold is calculated by

\begin{align}
    \label{eq:tracking}
    { T_i \in { T_1, T_2, ..., T_n } : \frac{\vert T_i - T_{ref} \vert}{T_{ref}} \leq 0.01 }
\end{align}

where $T_{ref}$ is the reference temperature and $T_i$ is each temperature measurement per part and the the rise time is calculated as

\begin{align}
    \label{eq:rise_time}
    t_{\mathrm{rise}} = \arg\min_k \left\{k: \frac{{\big| T_k - T_{ref}\big|}}{T_{ref}} \leq0.01\right \}
\end{align}

and represents how quicklz the controller can increase the ILT to within 1\% of the reference temperature.

\begin{table}
    \centering
    \caption{Number of layers with <1\% error values, rise time, heating and cooling modes, and total IDT duration for the entire builds.}
    \label{mean_increase_table}
    \begin{tabular}{@{}lccc@{}}
    \toprule
                     & Unsupported & Supported \\ \cmidrule(lr){1-4}
    \\[-1em]
    \textless{}1\% error {[}Layers{]} & 224 & 167 \\
    \\[-0.6em]
    Rise time {[}Layers{]} & 117 & 183 \\
    \\[-0.6em]
    Heating mode {[}Layers{]}    & 97 & 215    \\ 
    \\[-0.6em]
    Cooling mode {[}Layers{]}    & 186 & 68    \\ 
    \\[-0.6em]
    Total IDT duration {[}Minutes{]}    & 49.27 & 17.37   &  \\ 
    \\[-0.6em]
    \bottomrule
    \end{tabular}
    \end{table}

The unsupported part's ILT reaches the reference temperature at layer 117 and the controller switch initiates the cooling mode at layer 147, while the supported one reaches the ILT at layer 183 and the heating controller stays active until layer 265.
The controller decreases the laser power input from saturation to the higher process window to the lower as the overheating continues by effectively tracking the reference temperature.
The decay in laser power starts immediately after reaching the reference temperature and continues until reaching the lower boundary and initiating the cooling mode.
It takes 32 layers for the laser power to reach the lower boundary in the unsupported part while 82 layers in the supported part.
Although the inclusion of support structures increases the scanned area and thus the scanning time, the resulting increase in IDT (as shown in the second row of \Cref{fig:supported_unsupported_comparison} within the heating-controlled regions) is not significant.

%% file: 5-discussion.tex
\section{Discussion}
\label{sec:discussion}

\subsection{Switch-controller behavior}
\label{sec:switch_controller_behavior}

The switch controller for the diverging-converging cross-sectioned experimental part starts with the heating mode at layer 100 since the reference temperature is greater than the current temperature.
As per \Cref{fig:totem_plot} and \Cref{fig:comparison_plot}, ILT of the controlled part is observed to increase more rapidly in the first heating mode relative to the uncontrolled part.
This is due to the ILT being below the reference temperature, which causes the controller to request greater energy input.
In this case, the commanded power saturates at the upper bound of the process window.
Upon reaching the reference temperature, the power input is reduced and the reference temperature is tracked with a slight initial oscillation prior to settling.
The increasing cross-section causes the process to overheat, which causes the controller to reduce the commanded power until it is saturated to the lower end of the defined process window, as explained in detail in \Cref{sec:heat_buildup}.
Reaching the lower end triggers the controller switch signal to switch to the cooling mode as designed and depicted in \Cref{fig:proposed_method}.
The cooling controller is then initiated and starts adding dwell times to the start of each layer to allow for the ILT to return to the reference temperature.

The controlled part's ILT tracks the reference temperature with very low error in the cooling mode.
This was expected, as the dwell time is determined by frequent temperature measurements without any model or prediction, which turns out to be a very practical, efficient, and robust method.
Therefore, the triggering of the next layer sets the ILT deterministically. 
In the heating mode, the behavior of the laser power assignment and the ILT response is very similar to the findings of \cite{kavas2023layer}.
The controller proposed in this study is a model-based feedback optimization scheme that yields a structure analogous to a discrete-time PI controller, where proportional-like action stems from instantaneous temperature deviations and integral-like action from the accumulation of inter-layer deviations. 
In contrast to conventional PI/PID approaches, it relies on a simple data-driven thermal model rather than empirically tuned gains. 
By explicitly exploiting model information, the controller adapts to the process dynamics, improves performance over purely model-free PI formulations, and removes the need for manual gain tuning.
As it is stated in~\Cref{sec:implementation}, the fact that the employed linear controller model, identified on a single diverging angle, performs well in stabilizing ILT for both diverging and, more importantly, converging parts highlights the general robustness of the controller approach.

As shown in \Cref{fig:comparison_plot}, the heating-controlled and the uncontrolled parts show a very similar trend in ILT, with greater amounts of overheating observed for the uncontrolled part.
The peak temperature reached by the end of the diverging cross-section region is only lower in the heating-controlled part due to the effect of thermal history from the initial layers after the controller initiation.
In a subsequent trial, the part was built with a static $15-s$ dwell time. Similar to the results of \cite{nahr2025advanced} where a static $40-s$ IDT was necessary to print overhanging geometries in MS1, this indicates the utility of adding static dwell times to prevent overheating. 
However, the addition of static dwell times has two shortcomings over the proposed adaptive IDT assignment.
On one hand, the static dwell time is applied globally regardless of the actual thermal state, resulting in lost productivity when additional cooling is not necessary.
On the other, the assigned dwell time may be insufficient to maintain stability if the conditions leading to overheating are maintained over an extended period or made worse, e.g., by the geometry.

Another significant effect of overheating is observed in the corrosion patterns on the downward-facing edges of all the printed parts as shown in \Cref{fig:totem_buttom}.
The uncontrolled part shows extensive signs of discoloration, particularly in the corners (region b), as a result of remaining at elevated temperatures for an extended period. 
The corresponding regions are less pronounced in the heating-controlled and switch-controlled parts.
The highly discolored region immediately after the controller initiation in the laser power-controlled part shows the effect of the rise-time of the controller where the laser power input is saturated to the upper threshold until the ILT reaches the reference temperature.
It should be interpreted as the effect of greater energy input after the laser exposure, as opposed to the ILT, which is the measurement immediately before the exposure of the next layer. 
Overall, although the reference temperature is the same for all parts, the switch-controlled part exhibits the most uniform coloration profile. 
This indicates that the peak temperatures (which are not measured in the experimental procedure) during the layer-to-layer laser exposure modes are also reduced by the switch controller application.

The microstructural comparison shown in \Cref{fig:microstructure} indicates that the proposed switch control strategy preserves a stable and consistent microstructure for the investigated material and process parameter combination. 
Only an insignificant number of lack-of-fusion pores are observed in both the controlled and uncontrolled samples, with no systematic increase associated with the application of control. 
Compared to the uncontrolled case, the influence of regions close to the overhang is reduced in the switch-controlled part, resulting in a more uniform microstructural appearance in these geometrically sensitive areas.

Locally increased grain sizes spanning multiple layers are observed in regions where the laser power is increased to maintain the reference ILT. 
This behavior is expected, as elevated thermal input and reduced cooling rates promote grain coarsening, and is consistent with observations reported in the literature~\cite{kavas2023layer}.
Applying the proposed control strategy to other alloy systems or build configurations with stronger thermal sensitivity is expected to lead to more pronounced microstructural effects.
Overall, these results suggest that the proposed control approach effectively stabilizes the layer-to-layer thermal state while maintaining microstructural integrity and avoiding adverse defect formation.

\subsubsection{Symmetry}

In the \Cref{fig:totem_plot}, the peak dwell time in the controlled part and the peak ILT in the uncontrolled part are reached precisely where the cross-section changes from diverging to converging.
Similarly, the ILT changes from decreasing to increasing, where the cross-section changes from converging to diverging.
The ILT and dwell time trends shift without any lag relative to the layers, suggesting that the cross-section change trend per layer is dominant in defining the ILT against the thermal history of the part.

An interesting observation can be made regarding the thermal behavior of the converging 120° section (layers 633-833).
Here, the section begins with decreasing dwell time with a less steep slope compared to the other sections since the $\Delta a$ is the lowest.
Around layer 750, the controller switches to the heating mode since the cooling ILT reaches the reference temperature.
Here, the ILT tracks the reference by dynamically varying the laser power for the next 52 layers before the controller saturates at the upper limit of the process window.
The ILT continues to drop according to the cross-section change until layer 833, where the diverging section starts, and heat build-up again becomes dominant.

The asymmetric trajectory of laser power assignment is centered on the converging-diverging cross-section transition as observed in the \Cref{fig:totem_plot}.
The converging section has a flatter angle (120°) than the steep overhang angle (48°) of the following diverging section.
Due to the difference, the laser power is rapidly saturated to the lower threshold by the controller, hence the cooling mode is initiated.
This behavior exemplifies the laser power controller's decreasing effectiveness with steeper overhang angles.

\subsection{ILT and IDT estimation}
\label{sec:ilt_idt_estimation}
The proposed analytical approach for estimating ILT through equivalent conduction parameter fitting captures both the trend and the magnitude of ILT evolution remarkably well, particularly given its simplicity. 
In the first plot in~\Cref{fig:idt_prediction}, the calculated $\mathcal{K}_k$
Values qualitatively reflect variations in cooling rates across the entire build.
The cooling rate responds almost without layer lag to diverging and converging transitions, except at the initial change. 
Interestingly, $\mathcal{K}_k$ continues to decrease even as the cross-section and ILT begin to fall, a behavior driven by residual platform effects (i.e., a large thermal mass). As such, the fitted $\mathcal{K}_k$ values tend to underestimate the magnitude of conductive cooling. 

Since parameter fitting was performed only on a single overhanging section, the close agreement between the two models, especially in ILT prediction, indicates that layer-dependent conduction fitting is not essential for capturing temperature evolution or gross IDT trends through the printed geometry. 
Simple layer-wise area modeling is sufficient to reproduce both the temperature trajectory and its quantitative range.
Analytical approaches that describe multiple physical phenomena using lumped-parameter formulations have also been reported in the directed energy deposition literature~\cite{li2017extended,ansari2021analytical}. These models explicitly account for the moving heat source, whereas in the present study, the dominant thermal conduction is sufficiently captured using a one-dimensional formulation through the conduction term $\mathcal{K}_k$.
Moreover, the fact that a linear heating-mode controller successfully regulates ILT across both diverging and converging geometries demonstrates that effective ILT control can be achieved without complex 3D modeling. 
Contrary to previous studies on process control and observability~\cite{wood2023controllability}, our results suggest that full 3D temperature state estimation may not be required for closed-loop control of exposure-surface temperature within certain geometrical scales.

The predicted total waiting time (area under the prediction curves in the IDT prediction plot of~\Cref{fig:idt_prediction}) shows that predictions with constant and inverse proportional fit both add a total of 6 hours and 15 minutes, while the actual IDT addition in the controlled build is 4 hours 48 minutes.
IDT predictions exhibit a significant deviation from the controlled part, which is expected since the effects of strongly varying IDTs are not accounted for in the parameter estimation performed with a constant IDT.
The proposed model may be insufficient to capture the cooling trend of each within-layer cooling period, although it can capture the ILT trend when the IDT is fixed.
Since this study has been performed with a temperature-controlled build plate, further inaccuracy could be introduced in the case of builds with significant build plate temperature changes. Furthermore, the simplicity of this model and approach, for example, the temperature-independent material modeling, manual element height selection, and neglecting the part-scale thermal distribution, might limit the transferability to different materials, geometries, and machines without refitting.
Nevertheless, the resulting IDT trend can inform build planning by assigning additional exposure to different thermal bodies according to the per-layer IDT requirement. 
This enables a layer-specific IDT allocation strategy, formulated as an optimization problem, without resorting to full volumetric thermal field simulations. 
Such a feed-forward element, when combined with build-to-build and closed-loop control approaches, provides a promising direction for advanced LPBF process control.

\subsection{Trade-offs: Adding support structures versus dwell time}

The ILT of the unsupported inverted pyramid reaches the reference temperature 68 layers before the supported part (\Cref{fig:supported_unsupported_comparison}), indicating the reduced conduction area to the build plate.
Supporting the slower rise time finding, the ILT trend is steeper and the overshoot is more pronounced in the unsupported part.
Furthermore, laser power modulation remains within the allowable range for 32 layers for the unsupported part while the supported part's laser power is dynamically controlled for 82 layers before saturation and cooling mode onset.
Both of these findings are a clear result of the decreased overhanging rate due to the effect of the support structures.
Besides the heating controller behavior, the cooling mode in the unsupported part initiates earlier, and IDT assignment follows a steeper trend compared to the supported part.
For the entire build job, the unsupported part required a total of 49 minutes and 27 seconds of IDT assignment, while the supported part required 6 minutes and 32 seconds of IDT assignment.

While this means a significant increase in build time efficiency by support structure application, a much lower rise time in the unsupported part represents a trade-off for the overall build time.
An application where the reduced overheating is the main objective such as a distortion-prone geometry, support structure application favors the objective and efficiency more than applying a cooling controller.
However, for obtaining a stabilized ILT for an extended range of layers, the unsupported part with the proposed switch controller performs better with the expanse of increased build time.

Another significant finding is the 50\% increase in total energy consumption when building the same geometry with support structures. 
Although the choice of reference temperature plays an important role, the extended duration of the heating mode—coupled with the increased exposure area due to the support volumes—significantly contributes to this higher energy usage. 
Consequently, the benefits of shorter IDT assignments and reduced build time are offset by delays in tracking the reference temperature and increased energy consumption throughout the entire build process.
To repeat an obvious downside, supports must also be removed during post-processing, which also consumes valuable time, labor, and energy.

The proposed controller on an unsupported part outperforms the supported part in terms of achieving and maintaining the desired process conditions as quickly as possible.
Given that sustainability and efficiency are key value propositions of the PBF-LB/M process~\cite{nyamekye2024sustainability}, optimizing the interaction between ILT, energy efficiency, build time, and part quality remains a promising direction for future research.

The two possible approaches are described in \Cref{sec:strategies} to stabilize ILT: Reduction of the overhanging rate by support structure addition or a heating and cooling-based thermal process control.
As it is previously discussed in the work of Olleak et al. \cite{olleak2024understanding}, IDT addition per layer can be absorbed per part once there are multiple parts planned in the same build job.
Therefore, the results show that given a reference temperature and a fixed geometry, build time per part can only be reduced by adding more parts to the build plan so that the effective IDT per part is greater and the overheating rate is reduced.
Since such information can only be enabled via an accurate process model in a feed-forward manner, the details of this relationship are beyond the scope of this study and represent a valuable future research topic.

\subsection{Controller limitations}
\label{sec:limitations}

The switch controller proposed in this work extends the previously described~\cite{kavas2023layer} lower laser power range limit imposed by the process window.
By applying variable IDTs through closed-loop control, the overheating trend of the ILT can be stabilized indefinitely, depending on the range of the input parameters.
However, as described in the pre-converging to -diverging transition ILT trends in~\Cref{fig:totem_plot}, the cooling trend of the converging bodies may not be stabilized by the upper boundary saturated laser power.
This represents another limitation imposed by the higher end of the process window which prevents the meltpool to pass in the keyhole mode.
Similar to the lower limit, the upper limit is also highly dependent on the reference value selection.

Another practical implication of the study is the use of a temperature-controlled build platform.
Preheating the platform to 200 °C suppresses the early-build thermal transient and ensures that the fitted conduction parameter $K_k$ and the controller behavior are evaluated under a quasi-steady boundary condition. 
For the commercial PBF-LB/M systems that do not include active platform heating, the initial layers would be strongly influenced by the evolving thermal state of the substrate, and conductive heat loss to the build plate assembly would dominate during the early portion of the build. 
This would introduce a layer-dependent, nonstationary boundary condition that could degrade the quality of parameter identification and reduce the predictive accuracy of simple thermal models such as the one used here. 
Nevertheless, while the proposed methods do not rely on a heated build plate, the additional dynamics of the build plate heatup would need to be modeled in the ILT calculations.

\subsection{Future work on ILT stabilization}

This study covers an important research gap by evaluating the interplay between inherent cooling and heating mechanisms of the PBF-LB/M process.
First of all, for consolidating the taken approach further, predictive control is promising to enhance the tracking performance.
Beyond the investigated gap, the generalized approach for adjusting the energy input by laser power in the heating mode experimented can be extended for other defining parameters, such as the scan speed or the beam diameter.
Since the IDT additions yields in a longer process time due to the passive cooling, an active cooling apparatus can be implemented not to comprimize the total time when stabilizing the ILT.
In order to further compensate for the build time, adaptively defining the layer thickness as another means to modulate the energy input is regarded as a very promising and challenging possible next step due to possible secondary effects for layer thickness adjustments.

%% file: 6-conclusion.tex
\section{Conclusion}
\label{sec:conclusion}

In this work, ILT stabilization capabilities are explored for both converging and diverging geometries in the PBF-LB/M process.
A novel layer-to-layer closed-loop feedback controller design is proposed that is capable of stabilizing the interlayer temperature against overheating and cooling based on the cross-section variation across the layers.
Alternatively, the same controller is applied to a part with support structures, and the effect of the support structure is evaluated to the control capabilities of the ILT.
The controller is designed with a switch-based algorithm that adapts the IDT by ILT-based triggering of layer execution for cooling and laser power input modulation within the allowable process window.
The proposed method represents the first example in the literature that controls excessive cooling and overheating phenomena that are naturally present in the PBF-LB/M process.
Furthermore, a very simple analytical model to capture the exposure surface temperature is proposed with a conduction parameter value to be fit for the process conditions.
It is used to estimate the temperature progression as well as the IDT needed to reach down to the reference temperature based on the area progression.
The main findings of the study are:

\begin{itemize}
    \item The switched controller stabilizes the ILT regardless of the amount of overheating introduced by the geometry of the part by adding interlayer dwell time per layer,
    \item The controller employed in the heating mode is capable of tracking the reference temperature in a converging cross-sectional area despite the controller model having been identified on a diverging part, showing the robustness of the approach.
    \item The addition of support structures increases the heat flow from the exposure area to the build plate, reducing the rate of overheating. For the same reference temperature, less additional dwell time is required, while reaching the target temperature requires more layers compared to an unsupported part. A significant (50\%) increase in the total energy consumption was required to build the chosen geometry with support structures versus without.
    \item The simple analytical model is capable of estimating the ILT and resulting IDT well, demonstrating a minimal effect of the changing geometry on the cooling rate.
    
\end{itemize}

Future work could address how build orientation, geometry, and layout influence the time required to stabilize overheating, complementing the closed-loop control of ILT. 
In particular, evaluating the controller’s robustness with more complex geometries would provide valuable insight. 
More broadly, studies of cooling behavior in fixed geometries offer a promising direction for understanding ILT evolution within a layer’s time frame and for ensuring consistent part quality across different builds.